\documentclass[lettersize,journal]{IEEEtran}
\usepackage{amsmath,amsfonts}
\usepackage{algorithmic}
\usepackage{algorithm}
\usepackage{array}
\usepackage[caption=false,font=normalsize,labelfont=sf,textfont=sf]{subfig}
\usepackage{textcomp}
\usepackage{stfloats}
\usepackage{url}
\usepackage{verbatim}
\usepackage{graphicx}
\usepackage{cite}
\usepackage{epstopdf}
\usepackage{multirow} 
\usepackage{booktabs}
\usepackage[table,xcdraw]{xcolor}

\usepackage{xcolor}

\begin{document}

\title{UBiGTLoc: A Unified BiLSTM-Graph Transformer Localization  Framework for IoT Sensor Networks}

\author{Ayesh~Abu~Lehyeh, Anastassia~Gharib~\IEEEmembership{Member,~IEEE}, Tian~Xia~\IEEEmembership{Member,~IEEE}, Dryver~Huston, Safwan~Wshah~\IEEEmembership{Member,~IEEE}
\thanks{Ayesh~Abu~Lehyeh and Safwan~Wshah are with the Department of Computer Science, The University of Vermont, Burlington, VT 05405, USA (emails: Ayesh.Abulehyeh@uvm.edu, Safwan.Wshah@uvm.edu).}
\thanks{Anastassia Gharib  is with the Department of Computer Science and Engineering, American University of Sharjah, Sharjah 26666, UAE (email: AGharib@aus.edu).}
\thanks{Tian~Xia is with the Department of Electrical Engineering, The University of Vermont, Burlington, VT 05405, USA (email: TXia@uvm.edu).}
\thanks{Dryver~Huston is with the Department of Mechanical Engineering, The University of Vermont, Burlington, VT 05405, USA (email: Dryver.Huston@uvm.edu).}
\thanks{This research has been supported by the U.S. National Science Foundation with grant 2119485.}
\thanks{\textit{Corresponding author: Safwan~Wshah.}}}




\maketitle

\begin{abstract}
Sensor nodes' localization in wireless Internet of Things (IoT) sensor networks is crucial for the effective  operation of diverse applications, such as smart cities and smart agriculture. 
Existing sensor nodes' localization approaches heavily rely on anchor nodes within wireless sensor networks (WSNs). Anchor nodes are sensor nodes equipped with global positioning system (GPS) receivers and thus, have known locations. These anchor nodes operate as references to localize other sensor nodes. However, the presence of anchor nodes may not always be feasible in real-world IoT scenarios. Additionally, localization accuracy can be compromised by fluctuations in Received Signal Strength Indicator (RSSI), particularly under non-line-of-sight (NLOS) conditions. To address these challenges, we propose UBiGTLoc, a Unified Bidirectional Long Short-Term Memory (BiLSTM)-Graph Transformer Localization framework. The proposed UBiGTLoc framework effectively localizes sensor nodes in both anchor-free and anchor-presence WSNs. The framework leverages BiLSTM networks to capture temporal variations in RSSI data and employs Graph Transformer layers to model spatial relationships between sensor nodes. Extensive simulations demonstrate that UBiGTLoc consistently outperforms existing methods and provides robust localization across both dense and sparse WSNs while relying solely on cost-effective RSSI data. 
\end{abstract}

\begin{IEEEkeywords}
Internet of Things (IoT), Bidirectional Long Short-Term Memory (BiLSTM), Graph Transformers, Outdoor Localization, Received Signal Strength Indicator (RSSI), Wireless Sensor Networks (WSNs).
\end{IEEEkeywords}

\section{Introduction}
\IEEEPARstart{W}{ireless} sensor networks (WSNs) are pivotal in the expanding Internet of Things (IoT), 
supporting a multitude of applications ranging from smart agriculture to smart cities~\cite{9990579}. Sensor nodes serve as the fundamental components of WSNs. Deployed in outdoor environments, these sensor nodes work together to enable large-scale sensing and communication. In WSNs for the IoT, sensor nodes are heterogeneous. They sense different data and support different technologies. Typically, these networks comprise of two types of sensor nodes, which are anchor sensor nodes and regular sensor nodes. Anchor sensor nodes, also refereed to as anchor nodes, are equipped with global positioning system (GPS) for precise sensor nodes' localization. In contrast, regular sensor nodes lack GPS and rely on anchor nodes for positioning. Therefore, anchor nodes are more powerful, with greater energy and computational resources, while regular sensor nodes are designed for lower cost and easier scalability of the WSN. 

In wireless IoT sensor networks, sensor nodes' localization is essential for the accurate functioning of location-based services and network management~\cite{asaad_comprehensive_2022}. For example, in smart agriculture, precise localization is used to monitor environmental parameters in large fields. Environmental monitoring in smart cities further requires accurate sensor nodes' location to effectively track changes in air quality, water levels, and weather conditions. 
Traditional sensor nodes' localization methods, such as those that rely on GPS, are often impractical for large-scale IoT deployments due to their expense and high energy consumption. Additionally, GPS inaccuracies are exacerbated by signal obstruction and multipath effects in urban environments, such as smart cities, and by poor coverage in rural areas, such as in smart agriculture settings~\cite{kagi_localization_2022}. This underscores the need for alternative sensor nodes' localization approaches that are not solely dependent on GPS.

Over the years, localization techniques have evolved significantly. Nowadays, they are primarily categorized into range-free and range-based approaches \cite{wang_dv-hop_2024}. Range-free methods rely on connectivity and proximity information of sensor nodes. These methods typically rely on anchor nodes for sensor nodes' localization. Although they are believed to be simpler and more energy-efficient than range-based methods, they often face accuracy issues in heterogeneous or anisotropic environments~\cite{asaad_comprehensive_2022}. 
This is because range-free methods depend on indirect distance measurements. For instance, in~\cite{wang_dv-hop_2024}, a range-free enhanced distance vector-hop (DV-Hop) algorithm is introduced, which refines distance estimation by integrating multinode and hop loss data. Another range-free localization approach in \cite{10684810} utilizes the triple-anchor centroid method with expected hop progress (EHP) weighting to enhance distance estimation to anchor nodes. These methods, while improving upon traditional range-free limitations, still grapples with the inherent imprecision of indirect measurements, particularly in heterogeneous network deployments. Hence, range-free localization methods are not the best choice for the IoT.  

In contrast, range-based localization methods, which utilize distance or angle measurements like Received Signal Strength Indicator (RSSI), Time of Arrival (ToA), and Angle of Arrival (AoA), generally offer higher accuracy~\cite{Review}. 
Among these, RSSI-based methods stand out due to their cost-effectiveness and energy efficiency. For example, a sensor nodes' localization approach using RSSI is proposed  in~\cite{Behra_2021}. 
This method employs Similarity-Based Ricci Flow Embedding (SBRFE) to reduce non-Euclidean artifacts in RSSI-based distance estimates. It enhances localization accuracy without the need for expensive hardware required in ToA and AoA-based schemes. Similarly, a cost-effective RSSI-based localization approach is introduced in \cite{10579925}, utilizing only four fixed anchor nodes and an improved cuckoo search algorithm to optimize location estimates. Therefore, RSSI-based methods 
are becoming a promising alternative to GPS-based methods for sensor nodes' localization in wireless IoT sensor networks. 

Meanwhile, recent advancements in machine learning (ML), specifically in Graph Neural Networks (GNNs), have shown considerable potential in overcoming the challenges of sensor nodes' localization in WSNs \cite{yan_graph_2021}. This is because GNNs can effectively capture the complex spatial relationships between sensor nodes. Unlike traditional Convolutional Neural Networks (CNNs), which operate on regular grid-like structures, GNNs are specifically designed to process graph-structured data, making them highly suitable for WSN localization tasks~\cite{xu_self-attention_2024}.
In the context of GNNs, node representations are learned through a series of message-passing steps, where each node aggregates information from its neighbors to update its own representation.   However, a common limitation of existing GNN-based sensor nodes' localization approaches is their dependence on anchor nodes \cite{yan_graph_2021,yan_attentional_2023,zhang_accurate_2024}. 
This reliance on GPS-enabled anchor nodes can be problematic in real-world applications, where deploying and maintaining such sensor nodes is often impractical or cost-prohibitive.
Additionally, the dynamic nature of many environmental conditions (including noise and variability in signal propagation) cause the RSSI information to vary widely over short periods. Traditional GNN models often struggle to capture these rapid temporal changes, which are crucial for accurate localization in such fluctuating conditions. 

In this article, we propose UBiGTLoc, a Unified Bidirectional Long-Short-Term Memory (BiLSTM)-Graph Transformer Localization framework for wireless IoT sensor networks. UBiGTLoc leverages Bidirectional Long-Short-Term Memory (BiLSTM) networks to capture temporal variations in RSSI data, while Graph Transformers 
enhances the framework’s ability to focus on important sensor node interactions. This combination makes UBiGTLoc well-suited for complex localization tasks in dynamic and large-scale wireless IoT sensor networks. 
Compared to existing sensor nodes' localization schemes, our approach not only enhances the accuracy of localization 
but also diminishes the dependency on anchor nodes. 
The proposed UBiGTLoc framework is designed to adapt seamlessly to the presence or absence of anchor nodes, making it more versatile and resource-efficient solution for a broader array of implementations in different IoT scenarios.

\subsection{Related Work}

In recent years, various ML approaches have been increasingly applied to sensor nodes' localization in WSNs. They tend to address the limitations of traditional methods, such as the reliance on precise hardware and susceptibility to various environmental factors (i.e., non-line-of-sight (NLOS) conditions and signal interference). For instance,  support-vector machines (SVM) are used together with trilateration for RSSI fingerprinting-based sensor nodes' localization in Long Range (LoRa) networks in~\cite{Behra_2021}.  However, 
this method requires the presence of a fixed number of gateways, which act as anchor nodes to estimate distances between themselves and end devices (i.e., regular sensor nodes). 
In \cite{10680101}, a machine learning-based multilateration (ML-MTL) method is designed for LoRa-based WSN localization. By leveraging low-power relay systems to enhance RSSI accuracy, this approach improves distance estimation between anchor nodes and regular sensor nodes while maintaining energy efficiency.
In~\cite{kagi_localization_2022}, deep neural networks are utilized to optimize regular sensor nodes' localization accuracy through parametric analysis combining both RSSI and AoA information from anchor nodes. In~\cite{foliadis_reliable_2022}, reliable deep learning-based sensor nodes' localization using channel state information (CSI) fingerprints and multiple anchor nodes is investigated. A CNN is employed to enhance the localization accuracy under uncertainty conditions. Building on these advancements, a unified deep transfer learning model based on multi-layer perception (named as U-MLP) is proposed to handle sensor nodes' localization in diverse environments, both indoor and outdoor in~\cite{ahmed_unified_2024}. This model 
employs encoder-based transfer learning 
to train a single deep learning model capable of accurately predicting the localization of IoT sensor nodes across various  environments. Another deep learning-based approach is proposed in \cite{10828773}, where a DV-Hop localization method enhanced by whale optimization algorithm (WOA) and data augmentation improves accuracy by generating virtual anchor nodes around real anchor nodes. Though effective, these methods overlook the inherent graph structure of WSNs.

With the advent of GNNs, localization methods have advanced further, significantly improving scalability and localization accuracy. 
This is because in contrast to traditional ML techniques, graph-based approaches 
capture spatial relationships within data. 
Graph convolutional networks (GCNs) and attention-based graph neural networks (AGNNs) have been at the forefront of this progress. 
In~\cite{yan_graph_2021}, 
a GCN is utilized to address the challenges related to line-of-sight (LOS) and NLOS noise conditions. This work is then extended in~\cite{yan_attentional_2023}, where AGNNs  are employed for robust localization in massive WSNs. By leveraging attention mechanisms, the approach enhances accuracy through focusing on the most relevant information from numerous anchor nodes. Similarly, in~\cite{zhang_accurate_2024}, a combination of knowledge graphs and GNNs are employed to tackle the challenge of accurate localization in environments with both LOS and NLOS conditions. This method models communication data as knowledge graphs and applies a heterogeneous graph attention mechanism to infer unknown data representations. 
In~\cite{xu_self-attention_2024}, an approach for collaborative target tracking is proposed by integrating factor graph-based data fusion with GNN enhanced by self-attention mechanisms.  The self-attention mechanism allows the model to autonomously select and weigh complex network features. 
This model is suitable for handling multi-source heterogeneous data in multi-agent environments.


\label{sec:related_work}

\begin{table*}[t!]
\caption{Description of Symbols and Notations Used in the Article.}
\label{t}
\begin{center}
\begingroup
\renewcommand{\arraystretch}{1.5}
\begin{tabular}{llll}
\hline
\hline
\textbf{Symbol} & \textbf{Description} & \textbf{Symbol} & \textbf{Description} \\
\hline 
\hline
 $L$ & The length/width of a WSN field & $\alpha$ & Percentage of anchor nodes in a WSN \\
\hline
 $N$ & Total number of sensor nodes in a WSN & $N_a$ & The number of anchor nodes in a WSN \\
\hline
 $N_r$ & The number of regular sensor nodes in a WSN & $G$ & A graph representing a WSN \\
\hline
 $\mathcal{N}$ & Set of all sensor nodes in a WSN & $\mathcal{E}$ & A set of edges between sensor nodes in a WSN \\
\hline
 $\mathbf{A} = \{A_{ij}\}$ & Adjacency matrix with entries $A_{ij}$ & $d_{ij}$ & Distance between sensor nodes $i$ and $j$ \\
\hline
 $d_{\text{th}}$ & Distance threshold (i.e., radio range) & $(\hat{X}, \hat{Y})$ & Set of predicted $x$ and $y$ coordinates of sensor nodes \\
\hline
 $(x_i, y_i)$ & Ground truth coordinates of the $i$-th sensor node & $(\hat{x}_i, \hat{y}_i)$ & Predicted coordinates of the $i$-th sensor node \\
\hline
 $\text{RSSI}_{i,j}$ & RSSI at sensor node $i$ measured from sensor node $j$ & $n$ & Path loss exponent \\
\hline
 $\text{RSSI}_{o}$ & Received signal strength at a reference distance $d_o$ & $X_{\sigma}$ & Gaussian random variable with standard deviation $\sigma$ \\
 \hline
 $I$ & Gaussian interference term with standard deviation $\sigma_I$ & $\kappa$ & Interference scaling factor \\
\hline
 $\mathcal{N}_r$ & Set of regular sensor nodes in a WSN & $\mathcal{N}_a$ & Set of anchor nodes in a WSN \\
\hline
 $T$ & Time window size for measuring RSSI & $\mathbf{f}_i$ & Feature matrix of the $i$-th sensor node \\
\hline
 $\text{RSSI}_{i,j,t}$ & RSSI between sensor nodes $i$ and $j$ at time stamp $t$ & $\mathbf{\vec{f}}_{i,t}$ & Feature vector of the $i$-th sensor node at time stamp $t$ \\
\hline
 $\text{f}_{i,k,t}$ & Value of feature $k$ of the $i$-th sensor node at time $t$ & $\mathbf{F}$ & Collected features' matrix at the central unit \\
 \hline
 $N_n$ & Number of neighbors for the \(n\)-th sensor node & $h_n$ & Number of packets forwarded by \(n\)-th sensor node  \\
 \hline
 $C_n$ & Complexity for the \(n\)-th sensor node (anchor-free) & $C_n^r$ & Complexity for the \(n\)-th regular sensor node  \\
 \hline
 $C_n^a$ & Complexity for the \(n\)-th anchor sensor node & $C_{\text{total}}$ & Total system computational complexity  \\
\hline
 $\text{f}_{i,k,t}^{\prime}$ & Imputed value of feature $k$ for sensor node $i$ at time stamp $t$ & $\mathbf{F}'$ & Feature matrix after mean imputation \\
\hline
 $\tau$ & The number of non-missing feature values & $\mu_k$ & Mean value of feature $k$ \\
\hline
 $\sigma_k$ & Standard deviation of feature $k$ & $\mathbf{F}''$ & Feature matrix after Z-score normalization \\
\hline
$\text{f}_{i,k,t}^{\prime\prime}$ & Feature $k$ value after Z-score normalization at time stamp $t$
  & $\mathbf{F}''_t$ & Pre-processed 2D feature tensor at time $t$ \\
\hline
 $\mathbf{C}_t$ & Cell state of LSTM at time stamp $t$ & $\mathbf{h}_t$ & Hidden state of LSTM at time stamp $t$ \\
\hline
 $\mathbf{\lambda}_t$ & Forget gate of LSTM at time stamp $t$ & $\hat{\mathbf{C}}_t$ & Candidate cell state at time stamp $t$ \\
\hline
 $\mathbf{\phi}_t$ & Input gate of LSTM at time stamp $t$ & $\mathbf{o}_t$ & Output gate of LSTM at time stamp $t$ \\
\hline
 $\mathbf{W}_l$ & Learnable weight matrix at layer/module $l$ & $\mathbf{\vec{b}}_l$ & Learnable bias vector at layer/module $l$ \\
\hline
 $\overrightarrow{\mathbf{h}}_o$ & Final hidden state of BiLSTM (forward direction) & $\overleftarrow{\mathbf{h}}_o$ & Final hidden state of BiLSTM (backward direction) \\
\hline
 $\mathbf{g} = \{\mathbf{\vec{g}}_i\}$ & Temporal encoding output feature matrix with entries $\mathbf{\vec{g}}_i$ & $H_l$ & Hidden dimension of layer/module $l$ \\
\hline
  $\beta^e_{i,j}$ & Head $e$ attention coefficient between sensor nodes $i$ and $j$ &  $E$ & Total number of attention heads \\
\hline
 $\mathbf{\vec{Q}}_i^e$ & Query vector of sensor node $i$ for head $e$ & $\mathbf{\vec{K}}_j^e$ & Key vector of sensor node $j$ for head $e$ \\
\hline
 $\mathbf{\vec{V}}_j^e$ & Value vector of sensor node $j$ for head $e$ & $\mathbf{g}^\prime = \{\mathbf{\vec{g}}^\prime_i\}$ & TransformerConv output feature matrix with entries $\mathbf{\vec{g}}^\prime_i$ \\
\hline
 $\mathbf{g}^{\prime\prime}$ & Spatial attention output feature matrix & $\mu_B$ & Mean of a mini-batch in batch normalization \\
\hline
 $\sigma_B^2$ & Variance of a mini-batch in batch normalization & $\epsilon$ & Small constant to avoid division by zero \\
\hline
 $\hat{\mathbf{g}}^{\prime\prime}$ & Normalized output feature in batch normalization & $\mathbf{Z}$ & Batch normalization output feature matrix \\
\hline
 $\gamma$ & Scaling parameter in batch normalization & $\delta$ & Shifting parameter in batch normalization \\
\hline
\hline
\end{tabular}
\endgroup
\end{center}
\end{table*}

While these methods have significantly advanced sensor nodes'  localization in WSNs, they  underscore several limitations. Most existing approaches either rely heavily on the presence of anchor nodes or are tailored to specific network conditions, such as dense or sparse deployments. However, real-world  WSNs in IoT scenarios are often heterogeneous and may lack sufficient anchor nodes. Such environments may also face dynamic environmental changes that affect signal propagation and introduce temporal variations in signal strength. These variations are crucial for accurate localization, yet many methods overlook them. Additionally, many of these methods often rely on extra measurements (such as AoA) beyond cost-effective RSSI. These  require additional complex hardware and increase costs. Therefore, there is a need for a new solution that will address the challenges of these existing~works.

\subsection{Article Contributions}
 We propose UBiGTLoc, a unified sensor nodes' localization framework for wireless IoT sensor networks. 
 \textit{In contrast to the existing sensor nodes' localization methods, the contributions of this article are as follows}:
\begin{itemize}
\item We introduce a general solution that  effectively handles both anchor-free and anchor-presence WSNs, ensuring sensor nodes' localization in different IoT scenarios. 
\item Along with spatial information of the WSN graph structure, we  consider temporal  RSSI variations  and tackle NLOS conditions to  
improve localization accuracy in dynamic IoT environments. 
\item We rely solely on RSSI measurements, avoiding extra costy options (such as GPS, AoA, and ToA), to reduce the required sensor nodes' hardware complexity.
\item The proposed solution is designed to guarantee stable localization accuracy for various network sizes,  even with fewer number of sensor nodes present in the WSN field. 
\end{itemize}

\noindent UBiGTLoc operates  in a centralized manner  and every pre-defined period of time.  It consists of four main modules:
\begin{enumerate}
    \item In the first module, \textit{Data Pre-Processing}, we handle data imputation and normalization to ensure high-quality inputs. This module prepares the RSSI data for subsequent processing by addressing missing values and scaling features to maintain consistency.
    \item In the second module, \textit{Temporal Encoding}, we utilize BiLSTM to capture temporal dependencies in RSSI data. This approach mitigates the effects of noise and signal fluctuations and thus, enhances localization robustness under LOS and NLOS conditions.
    \item In the third module, \textit{Spatial Attention}, we employ two Graph Transformer layers 
    to model WSN's spatial dependencies. This module leverages the WSN's graph structure  and
    utilizes an attention mechanism to focus on the most relevant interactions between sensor nodes. 
    \item In the fourth module, \textit{Global Synthesis}, we use a fully connected layer to transform the high-dimensional data from the previous modules into the final predicted location of each sensor node. 
\end{enumerate}
\noindent We finally present simulation results to evaluate and compare the performance
of the proposed UBiGTLoc framework  to existing  localization methods. For this purpose, we simulate realistic NLOS conditions and RSSI fluctuations in dense and sparse WSNs with and without anchor~nodes. 

\begin{figure*}[t!]
    \centerline{\includegraphics[width = 0.96\textwidth]{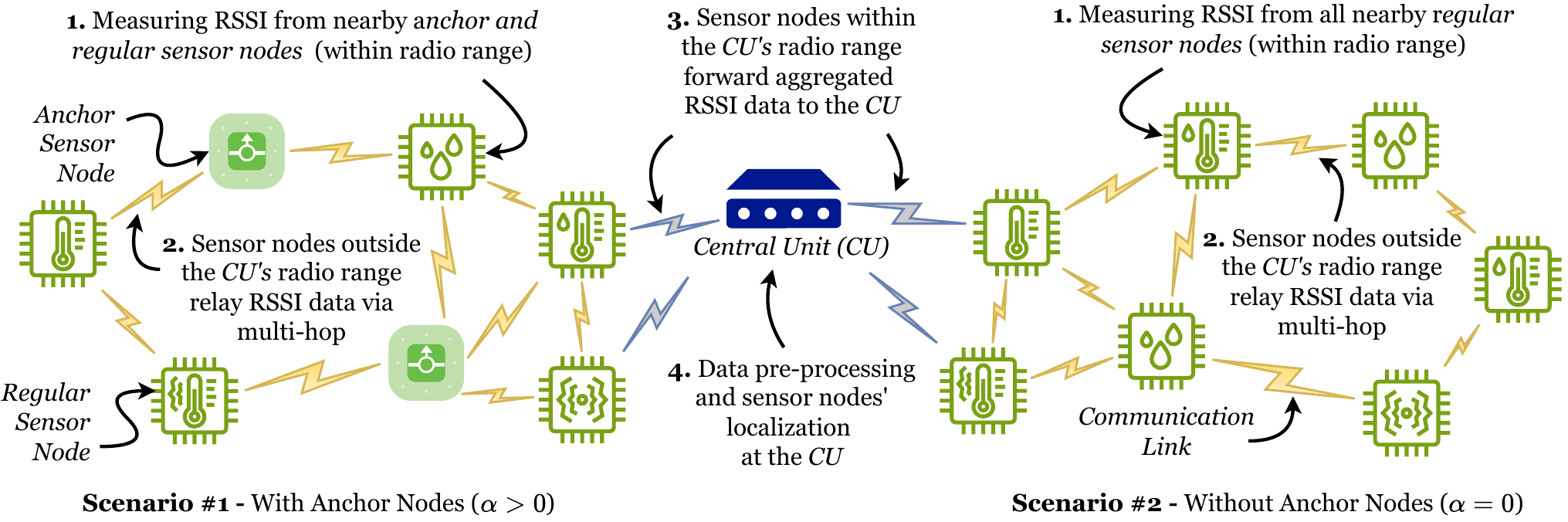}} 
      \caption{The system model of wireless IoT sensor networks with and without anchor nodes.}
      \label{systemmodel}
\end{figure*} 

\subsection{Article Organization}
The rest of the article is organized as follows. Section~\ref{sec:sysmodel} describes the proposed system model of a WSN with and without anchor nodes. The proposed UBiGTLoc framework is then presented in Section~\ref{sec:UBiGTLoc}. Section~\ref{sec:sim_settings} covers simulation settings and training procedure. Section~\ref{sec:sim_results} demonstrates simulation results. 
Section~\ref{sec:conclusion} concludes the article. 
Table~\ref{t} describes all the symbols and notations used in the~article.

\section{System Model and Problem Formulation} \label{sec:sysmodel}
Fig.~\ref{systemmodel} shows the proposed system model. We consider a heterogeneous WSN, where sensor nodes can be of various kinds, each with distinct capabilities and functionalities. These sensor nodes can be, for example,  part of a smart city network, equipped with sensors like environmental sensors, energy trackers, 
cameras, and intelligent streetlights~\cite{10258332}. Additionally, they can be integrated into 
machinery, supporting various urban and rural IoT applications. This approach allows us to model a wide range of IoT environments, accommodating sensors with different measurement accuracies, communication ranges, and processing powers.  Sensor nodes are initially randomly deployed within a square field area of \(L \times L~m^2\). However, they can experience drifts in their positions over time due to environmental factors such as wind or flooding. Hence, sensor nodes' localization is required to determine their actual location with time. In our system model, sensor nodes communicate within a limited radio range. Therefore, not every sensor node can directly communicate with the central unit that manages the underlying WSN. Instead, only sensor nodes within the radio range of the central unit send data directly, while other sensor nodes forward their data through intermediate sensor nodes. This approach reflects real-world WSN scenarios, where direct communication between all sensor nodes and the central unit is often impractical due to range limitations, interference, and energy constraints.

We consider two types of sensor nodes in the WSN. These are regular sensor nodes and anchor nodes. Anchor nodes are sensor nodes with 
known locations from initial deployment or equipped with GPS modules to determine their corresponding locations \cite{10537989}. They serve as reference points, thereby enabling accurate localization of regular sensor nodes within the WSN. Depending on the specifics of the WSN environment and an IoT application, the percentage of anchor nodes (i.e.,~$\alpha$) in a WSN may vary.  For instance, in smart cities, the infrastructure allows for the easy installation of anchor nodes. Hence, anchor nodes can be deployed extensively to ensure high localization accuracy across densely populated areas. Conversely, in rural areas with smart agriculture, GPS signals may be unreliable due to poor satellite coverage.  
Thus, anchor nodes might not always be feasible in such scenarios.

Fig.~\ref{systemmodel} shows two scenarios, where the percentage of anchor nodes in the WSN is greater than zero (i.e., $\alpha >0$) and where the percentage of anchor nodes is zero (i.e., $\alpha=0$). In the first scenario, anchor nodes constitute a fraction of the total number of sensor nodes, denoted as \(N_a = \alpha \times N\), where 
$N$ is the total number of sensor nodes in a WSN. Thus, the total number of sensor nodes is given by $N=N_a + N_r$, where $N_r$ denotes the number of regular sensor nodes. In WSNs with anchor nodes, the placement of anchor nodes can be random or based on a grid pattern, depending on the IoT application needs and the WSN environment~\cite{9716049}. Without loss of generality, in this article, we assume random placement of anchor nodes. 
In the second scenario, since the percentage of anchor nodes is zero (i.e., $\alpha =0$), $N_a = 0$ and $N = N_r$. In both scenarios and as mentioned previously, a  central unit is used to manage the underlying WSN. It acts as a sink node, which collects sensed data over time and assists in sensor nodes' localization. We  next go through the problem formulation for sensor nodes' localization that applies to both of the aforementioned scenarios, i.e., with and without anchor~nodes.


\subsection{Sensor Nodes' Localization Problem Formulation}
We formulate sensor nodes' localization in a WSN as a graph-based problem. 
The WSN is represented as a graph $G = (\mathcal{N}, \mathcal{E})$, where $\mathcal{N}$ is the set of sensor nodes, including both regular sensor nodes and anchor  nodes, and $\mathcal{E}$ represents the edges between sensor nodes based on their connectivity. This means that each \emph{i-th} sensor node  is represented by a vertex, i.e., sensor node \(i \in \mathcal{N}\), and the connectivity between sensor  nodes is captured using an adjacency matrix~\(\mathbf{A}\). The adjacency matrix~\(\mathbf{A}\) encodes the spatial relationships and the WSN graph structure. An entry \(A_{ij}\)  of the adjacency matrix~\(A\) is then  defined as:
\begin{equation}
A_{ij} =
\begin{cases}
1 & \text{if } d_{ij} \leq d_{\text{th}}, \\
0 & \text{otherwise},
\end{cases}
\end{equation}
\noindent where \(d_{ij}\) is the distance between sensor nodes \(i\) and \(j\) and \(d_{\text{th}}\) is the distance threshold that defines the radio range within which sensor nodes can communicate. 

We assume that each sensor node can estimate distance to its neighboring sensor nodes through 
the RSSI measurements.  Thus, each  sensor node is associated with RSSI measurements collected over time from the neighboring sensor nodes (the RSSI data collection process will be discussed in detail in Section~\ref{sec:rssi_coll}). These RSSI measurements capture both the temporal variations and the dynamic behavior of the WSN, and hence, can be used as features for our sensor nodes' localization problem.

The goal of the sensor nodes' localization problem is to estimate the positions of regular sensor nodes, i.e., $(\hat{X}, \hat{Y})$, based on the graph structure and the aforementioned RSSI measurements, where \(\hat{X} = \{\hat{x}_1, \hat{x}_2, \ldots, \hat{x}_{N_r}\}\) and \(\hat{Y} = \{\hat{y}_1, \hat{y}_2, \ldots, \hat{y}_{N_r}\}\) are the sets of predicted \(x\) and \(y\) coordinates, respectively. This is done by the central unit which  at first pre-processes the data (i.e., the adjacency matrix \(\mathbf{A}\) and the RSSI measurements made by 
sensor nodes') 
and then takes it as an input to a deep learning framework. 
Therefore, we formulate an optimization problem that minimizes the localization error of the predictions, which is the mean squared error (MSE). The formulated optimization problem is defined as follows:
\begin{equation} \label{opt1}
\left(\hat{X}, \hat{Y}\right)
= \arg \min_{\hat{X}, \hat{Y}} \frac{1}{N_{r}} \sum_{i=1}^{N_{r}} \left[ (x_i - \hat{x}_i)^2 + (y_i - \hat{y}_i)^2 \right],
\end{equation}
\noindent where \((x_i, y_i)\) is the ground truth coordinate of the \emph{i-th} regular sensor node, and \((\hat{x}_i, \hat{y}_i)\) is the predicted coordinate of the \emph{i-th} regular sensor node. 
 The  objective of the problem in Eq.~\eqref{opt1} is to estimate the locations of regular sensor nodes in a WSN, such that the predicted positions \((\hat{x}_i, \hat{y}_i)\) 
 are as close as possible to the true positions~\((x_i, y_i)\) for each \emph{i-th} regular sensor node. 
Following, we cover the utilized model to simulate  the RSSI measurements between sensor nodes.


\subsection{The Received
Signal Strength Indicator \& Network Density-Dependent Interference Model}
We employ the log-normal shadowing model to effectively capture the shadowing effects between sensor nodes~\cite{10609490}. To account for both environmental noise and network interference, the RSSI value at the \emph{i-th} sensor node measured from the \emph{j-th} sensor node (i.e., $\text{RSSI}_{i,j}$) is defined as follows:
\begin{equation}
\text{RSSI}_{i,j} = \text{RSSI}_{o} - 10n \log_{10} \left(\frac{d_{ij}}{d_o}\right)  - X_{\sigma} - I,
\label{eq:rssi_full}
\end{equation}
\noindent where $n$ is the path loss exponent, and $\text{RSSI}_{o}$ is the signal strength received from a reference distance $d_o$. The model includes two distinct stochastic terms. The first, $X_{\sigma}$, is a Gaussian random variable with zero mean and a standard deviation $\sigma$, which represents the combined effects of shadowing and NLOS conditions. 
The second term, $I$, is introduced to model the multi-access interference from simultaneous transmissions in a WSN.

There exist multiple interference models for WSNs~\cite{5462980}. In WSNs, interference typically arises from nearby sensor nodes and their overlapping communication ranges. In this work, we assume a Gaussian random process for the interference model, an approach that is effective for representing the combined effect of many independent signal sources~\cite{5549941}. 
This makes the Gaussian random process a great choice for modeling density-dependent interference. 
Additionally, the flexibility of Gaussian random processes allows for more realistic representations of the increasing signal contention found in denser networks~\cite{5549941}. Specifically, we assume that interference is modeled as a zero-mean Gaussian random variable with standard deviation $\sigma_I$. Inspired by the work in~\cite{interference}, we define the standard deviation $\sigma_I$ to be dependent on the number of sensor nodes $N$ as follows:
\begin{equation}
\sigma_I = \kappa \sqrt{N},
\label{eq:sigma_interference}
\end{equation}
\noindent where $\kappa$ is a scaling factor ($0 \leq \kappa \leq 1$) that reflects WSN  characteristics such as transmission power and communication range. Hence, larger values of $\kappa$ corresponds to higher interference impact, whereas $\kappa=0$ represents an ideal, interference-free environment. Similarly, as the number of sensor nodes in a WSN grows, the interference increases due to the larger number of simultaneous transmissions.

In real-world scenarios, RSSI values are directly measured at each sensor node. For simulation purposes, we assume that the initial  distances between sensor nodes are known. Using these known distances, we generate simulated RSSI values based on the Eq.~\eqref{eq:rssi_full}. This approach creates a controlled experimental environment to validate the proposed  methods. We next discuss the RSSI data acquisition processes used in this article.

\subsection{The RSSI Data Acquisition} \label{sec:rssi_coll}
Algorithm \ref{alg:UBiGTLoc} outlines the end-to-end process of collecting the RSSI data, i.e., from measuring it at sensor nodes to obtaining these readings by the central unit. As mentioned previously, we define \(\mathcal{N}\) as the set of sensor nodes, which is composed of both regular sensor nodes and anchor nodes. Specifically, \(\mathcal{N} = \mathcal{N}_r~\cup~\mathcal{N}_a\), where \(\mathcal{N}_r\) represents the subset of regular sensor nodes, and \(\mathcal{N}_a\) represents the subset of anchor nodes. The positions of all anchor nodes (if they are available in the WSN) are defines as $(x_a,y_a)$, where $a=\{1, \dots, N_a\}$. RSSI values are to be measured at each regular sensor node over the duration of a time window denoted as \(T\).  

\begin{algorithm}
\caption{The RSSI Data Acquisition}
\label{alg:UBiGTLoc}
\begin{algorithmic}[1]
\STATE \textbf{Input:} $\mathcal{N} = \{\mathcal{N}_r,\mathcal{N}_a\};~(x_a,y_a)~\forall~a=\{1, ..., N_a\};~T;~\alpha$   
\STATE \textbf{Output:} $\mathbf{F} \in \mathbb{R}^{N \times (N+2) \times T}$
\STATE \textbf{Step 1 -} Initialization
\STATE $\mathbf{f}_i~\leftarrow~0~\forall~\text{\emph{i-th} sensor node}~\in~\mathcal{N}$,
\STATE $\mathbf{F} \leftarrow 0$ at central unit
\STATE \textbf{Step 2 -} Sensor Nodes' Measurements 
\FOR{$t = \{1, \dots, T\}$}
    \FOR{$i = \{1, \dots, N\}$}
        \IF{$\alpha > 0$}  
            \IF{sensor node $i \in \mathcal{N}_r$}
                 \STATE Measure $\text{RSSI}_{i,j,t}~\forall~j=\{1, .., N\}$, where $i\neq~j$ 
                \STATE $\mathbf{\vec{f}}_{i,t} \leftarrow \text{RSSI}_{i,t}$ 
                
            \ELSE
                \STATE $\mathbf{\vec{f}}_{i,t} \leftarrow (x_i,y_i)=(x_a,y_a) $
            \ENDIF
        \ELSE
            \STATE Measure $\text{RSSI}_{i,j,t}~\forall~j=\{1, .., N_r\}$, where $i\neq j$ 
             \STATE $\mathbf{\vec{f}}_{i,t} \leftarrow \text{RSSI}_{i,t}$ 
        \ENDIF
    \ENDFOR
\ENDFOR
\STATE \textbf{Step 3 -} Central Unit Obtaining Measurements 
\STATE Update $\mathbf{F} \leftarrow 
\mathbf{f}_i~\forall~i=\{1, \dots, N\}$
\STATE \textbf{Step 4 -} Repeat Every Time Period $T$ 
\STATE GOTO \textbf{Step 2}
\end{algorithmic}
\end{algorithm}

The RSSI data acquisition process consists of four steps. At the first step, each \emph{i-th} sensor node initializes a feature matrix~$\textbf{f}_i$. 
Specifically, $\textbf{f}_i$ is defined as a matrix of feature vectors $\mathbf{\vec{f}}_{i,t}$ over time stamps $t = \{1, 2, \dots, T\}$~as:
\begin{equation}
\mathbf{f}_i = \left[\mathbf{\vec{f}}_{i,1}, \mathbf{\vec{f}}_{i,2}, \dots, \mathbf{\vec{f}}_{i,T}\right]^T,
\end{equation}
\noindent where the feature vector $\mathbf{\vec{f}}_{i,t}$ at any time stamp $t$ has a length of $N+2$ and is structured as follows:
\begin{equation} \label{eq1}
\mathbf{\vec{f}}_{i,t} = \left[\text{f}_{i,1,t}, \text{f}_{i,2,t}, \dots, \text{f}_{i,N+2,t}\right].
\end{equation}

\noindent In Eq.~\eqref{eq1}, $\text{f}_{i,k,t}$ represents a scalar value for each element in~$\mathbf{\vec{f}}_{i,t}$, where $k$ ranges from 1 to $N+2$. 
The first $N$ features, i.e., $\text{f}_{i,1,t}$ to $\text{f}_{i,N,t}$, 
correspond to the RSSI values $\text{RSSI}_{i,j,t}$ 
between the $i$-th and $j$-th sensor nodes in the network at time $t$, if measured in step two. Otherwise, their value will correspond to zero. 
The last two features, i.e.,  $\text{f}_{i,N+1,t}$ and $\text{f}_{i,N+2,t}$, represent the known coordinates $x_i$ and $y_i$ of the $i$-th sensor node, which are defined only for anchor nodes. This means that Eq.~\eqref{eq1} can be rewritten as follows:
\begin{equation} \label{eq2}
\mathbf{\vec{f}}_{i,t} = \left[\text{RSSI}_{i,1,t}, \text{RSSI}_{i,2,t}, \dots, \text{RSSI}_{i,N,t}, x_{i},y_{i}\right],
\end{equation}
\noindent where
\begin{align}
(x_i,y_i) =
\begin{cases}
(x_a,y_a), & \text{if sensor node $i \in \mathcal{N}_a$} \\
(0,0), & \text{otherwise}.
\end{cases}
\label{eq3}
\end{align}
\noindent The central unit also initializes a feature matrix $\mathbf{F}$ that eventually collects all features of the sensor nodes in the WSN (i.e.,~$\mathbf{f}_{i}~\forall~i~\in \mathcal{N}$). Thus, the dimensions of $\mathbf{F}$ are $N  \times (N+2) \times T$ (i.e., $\mathbf{F} \in \mathbb{R}^{N \times (N+2) \times T}$). 

In step two, the RSSI values are collected over multiple samples, i.e., $t = \{1,\dots, T\}$, within a time window $T$ to capture temporal variations at each sensor node. This means that during time period \(T\), multiple RSSI samples, i.e.,  \(\text{RSSI}_{i,j,t}\), are collected  for each pair of sensor nodes \((i, j)\), where \(t\) is a time stamp.  In accordance with the scenario, the RSSI data acquisition for each \emph{i-th} sensor node, where $i=\{1,2,...,N\}$, is as outlined~below.
\subsubsection{Scenario \#1 - With Anchor Nodes ($\alpha > 0$)} 
If anchor nodes are present in a WSN, Algorithm~\ref{alg:UBiGTLoc} first  checks whether a sensor node is an anchor node or not. If the \emph{i-th} sensor node is detected as a regular sensor node (i.e., sensor node~$i \in \mathcal{N}_r$), RSSI values (i.e., \(\text{RSSI}_{i,j,t}\)) are collected by 
this sensor node from each of its neighboring sensor node $j$ whether it is an anchor node (i.e., sensor node~$j \in \mathcal{N}_a$) or a regular sensor node (i.e., sensor node~$j \in \mathcal{N}_r$). 
These values are used to update the feature vector \(\mathbf{\vec{f}}_{i,t}\) at each \emph{i-th} regular sensor node.
Otherwise, if a sensor node is an anchor node, its feature \(\mathbf{\vec{f}}_{i,t}\) is set to its known position~\((x_i,y_i)=(x_a,y_a)\).
\subsubsection{Scenario \#2 - Without Anchor Nodes ($\alpha = 0$)} When anchor nodes are not present (i.e., $N=N_r$), RSSI values are collected from regular sensor nodes only. In this scenario, at each \emph{i-th} regular sensor node, RSSI values (i.e., \(\text{RSSI}_{i,j,t}\)) are measured from its neighboring regular sensor nodes, where $i\neq j$. 
Similarly, these values are used to update the feature vector \(\mathbf{\vec{f}}_{i,t}\) at each \emph{i-th} regular sensor node.

It is worth noting that each sensor node sends the beaconing signal periodically to allow neighboring sensor nodes to measure the RSSI value. Traditionally, such transmissions are sparse to minimize collisions and are done with limited transmission range to avoid interference~\cite{9985558, 9016059}. In addition, carrier-sense multiple access/collision avoidance (CSMA/CA) with randomized backoffs minimize concurrent transmissions and collisions, bounding interference to levels empirically validated in low-cost deployments (e.g., ESP32/Zigbee)~\cite{9985558, 9016059}. To identify the sources of beaconing signals, each sensor node is assigned a unique identifier (i.e., a 16-bit address) embedded in packet headers. When a sensor node receives a beacon from its neighboring sensor node, it decodes the sender’s identifier (ID) from the header and records the RSSI alongside this ID, maintaining a dynamic neighbor table. This lightweight process ensures that sensor nodes map RSSI measurements to specific neighbors without requiring synchronization or wideband signals, which align with low-cost constraints.

In step three, the collected features from each sensor node are then sent to the central unit. The sensor nodes within the central unit’s radio range send data directly, while sensor nodes not in the radio range of the central unit forward their data through intermediate sensor nodes, as shown in Figure 1. Upon receiving the data from sensor nodes, the central unit constructs a feature matrix \(\mathbf{F}\) for all sensor nodes 
as follows:
\begin{equation}
    \mathbf{F} =
\begin{pmatrix}
\mathbf{\vec{f}}_{1,1} & \mathbf{\vec{f}}_{1,2} & \dots & \mathbf{\vec{f}}_{1,T} \\
\mathbf{\vec{f}}_{2,1} & \mathbf{\vec{f}}_{2,2} & \dots & \mathbf{\vec{f}}_{2,T} \\
\vdots & \vdots & \ddots & \vdots \\
\mathbf{\vec{f}}_{N,1} & \mathbf{\vec{f}}_{N,2} & \dots & \mathbf{\vec{f}}_{N,T} \\
\end{pmatrix}
\end{equation}
\noindent It is worth noting that generally, if sensor nodes $i$ and $j$ are not in the radio range (i.e., \(A_{ij} = 0\)), the measured RSSI value is zero (i.e., \(\text{RSSI}_{i,j,t} = 0\)). Additionally, the RSSI value for self-connectivity \(\text{RSSI}_{i,i,t}\) is set to a value that approaches infinity (\(\text{RSSI}_{i,i,t} \rightarrow \infty\)) to signify strong self-connectivity for the \(i\)-th sensor node. Moreover, in the constructed feature matrix \(\mathbf{F}\), the features for anchor nodes will be zeros except for the last two features representing the $x_a$ and $y_a$ values. Conversely, for regular sensor nodes, the last two features corresponding to \(x\) and \(y\) coordinates will be zeros, while the RSSI values will be non-zero (if within the radio range).

As denoted by step four, the process of harvesting the RSSI data by sensor nodes and obtaining it by the central unit is repeated over again. Meanwhile, the matrix \(\mathbf{F}\), composed of all \(\mathbf{\vec{f}}_{i,t}\) vectors from all sensor nodes, is  well-structured and formatted for further localization processing at the central unit. This comprehensive representation allows the system to account for various network conditions, including connectivity, range limitations, and the presence or absence of anchor nodes. In the following section, we discuss the proposed UBiGTLoc framework for sensor nodes' localization in WSNs.

\subsection{Complexity Comparison and Analysis of Anchor-Free and Anchor-Based Sensor Nodes Localization Scenarios}
In this section, we analyze and compare the computational complexity of each system component in both scenarios (i.e., anchor-based and anchor-free sensor nodes' localization). Let \( N_n \) be the number of neighbors for the \emph{n-th} sensor node (where \( N_n \ll N \) due to the limited radio range of sensor nodes), and let \( h_n \) denote the number of packets that sensor node \(n\) must transmit or forward (including its own and possibly other sensor nodes’ data) to the next hop.
\subsubsection{Anchor-Free Scenario}
In the anchor-free scenario, each sensor node measures the RSSI from each of its neighbors over \( T \) timestamps. Since measuring RSSI from a single neighbor for \( T \) timestamps is an \(\mathcal{O}(T)\) operation, the total measurement complexity per sensor node is $\mathcal{O}(T \cdot N_n)$. 
After measuring RSSI values, each \emph{n-th} sensor node transmits \( h_n \) packets to the next hop, leading to a transmission complexity of $\mathcal{O}(h_n)$ for each \emph{n-th} sensor node. 
Thus, the overall per-sensor node complexity in the anchor-free scenario is:
\begin{equation}
    C_{n}  = \mathcal{O}(T \cdot N_n + h_n).
\end{equation}
For all $N$ sensor nodes, the total complexity of the localization process becomes: 
\begin{equation}
C_{\text{total}} 
= \mathcal{O}\Bigl(\sum_{n=1}^{N} \bigl(T \cdot N_n + h_n\bigr)\Bigr).
\end{equation}
Notably, this scenario requires no specialized hardware and maintains relatively balanced overhead distribution. 
\subsubsection{Anchor-Based Scenario}
In the anchor-based scenario, regular sensor nodes perform the same RSSI measurements and transmissions as in the anchor-free case, maintaining the same per-sensor node complexity:
\begin{equation}
    C_{n}^{r} = \mathcal{O}(T \cdot N_n + h_n).
\end{equation}
Anchor nodes, however, do not measure RSSI. Instead, they acquire their location \((x_a, y_a)\) using GPS at each timestamp, resulting in a location identification  complexity of $\mathcal{O}(T)$ for each anchor node.
The transmission complexity for an anchor node, considering it relays \( h_n \) packets to the next hop, is $\mathcal{O}(h_n)$. 
Thus, the total complexity for an anchor node is:
\begin{equation}
    C_n^{a} = \mathcal{O}(T + h_n).
\end{equation}
\noindent Therefore, for all $N = N_r +N_a$ sensor nodes, the total complexity of the localization
process becomes:
\begin{equation} \small
\begin{aligned}
    C_{\text{total}} 
    & = \mathcal{O}\Bigl(\sum_{n=1}^{N_r} \bigl(T \cdot N_n + h_n\bigr) + \sum_{n=N_r+1}^{N} \bigl(T + h_n\bigr)\Bigr) \\ & = \mathcal{O}\Bigl(T \cdot \sum_{n=1}^{N_r} N_n + \sum_{n=1}^{N_r} h_n + T \cdot N_a + \sum_{n=N_r+1}^{N} h_n \Bigr)
    \\&= \mathcal{O}\Bigl(
    T \cdot \sum_{n=1}^{N} N_n 
    + \sum_{n=1}^{N} h_n 
    - T \cdot \Bigl( \sum_{n=N-N_a+1}^{N} N_n - N_a
    \Bigr)
    \Bigr).
\end{aligned}
\end{equation}

In summary, the anchor-free scenario offers the advantage of requiring no additional hardware with the computational complexity more evenly distributed across the sensor nodes than in the anchor-based scenario. This results in a more balanced energy consumption profile, which is crucial for the overall longevity of WSNs. Even though the anchor-based scenario reduces the overall computational burden by $\mathcal{O}\bigl(T \cdot \bigl( \sum_{n=N-N_a+1}^{N} N_n - N_a\bigr)\bigr)$, it introduces additional hardware requirements (i.e., GPS) and thus higher energy consumption. This is because GPS measurements require more energy consumption than RSSI measurements \cite{8614863}. It thus results in non-uniform operations between regular and anchor sensor nodes, leading to unfair energy expenditure between them. In addition, since the number of anchor nodes ($N_a$) is typically much smaller than the number of regular sensor nodes ($N_r$), any advantage gained by reducing the overall complexity in the anchor-based scenario can become negligible for very large WSNs. This is because $N$ grows while $N_a$ remains small. Hence, the term $\mathcal{O}\bigl(T \cdot \bigl( \sum_{n=N-N_a+1}^{N} N_n - N_a\bigr)\bigr)$ becomes negligible, effectively converging to the same overall complexity as the anchor-free scenario. Still, in both scenarios, sensor nodes closer to the central unit experience higher transmission burdens due to their relaying~duties. 


\begin{figure*}[t!]
\centerline{\includegraphics[width =1 \textwidth]{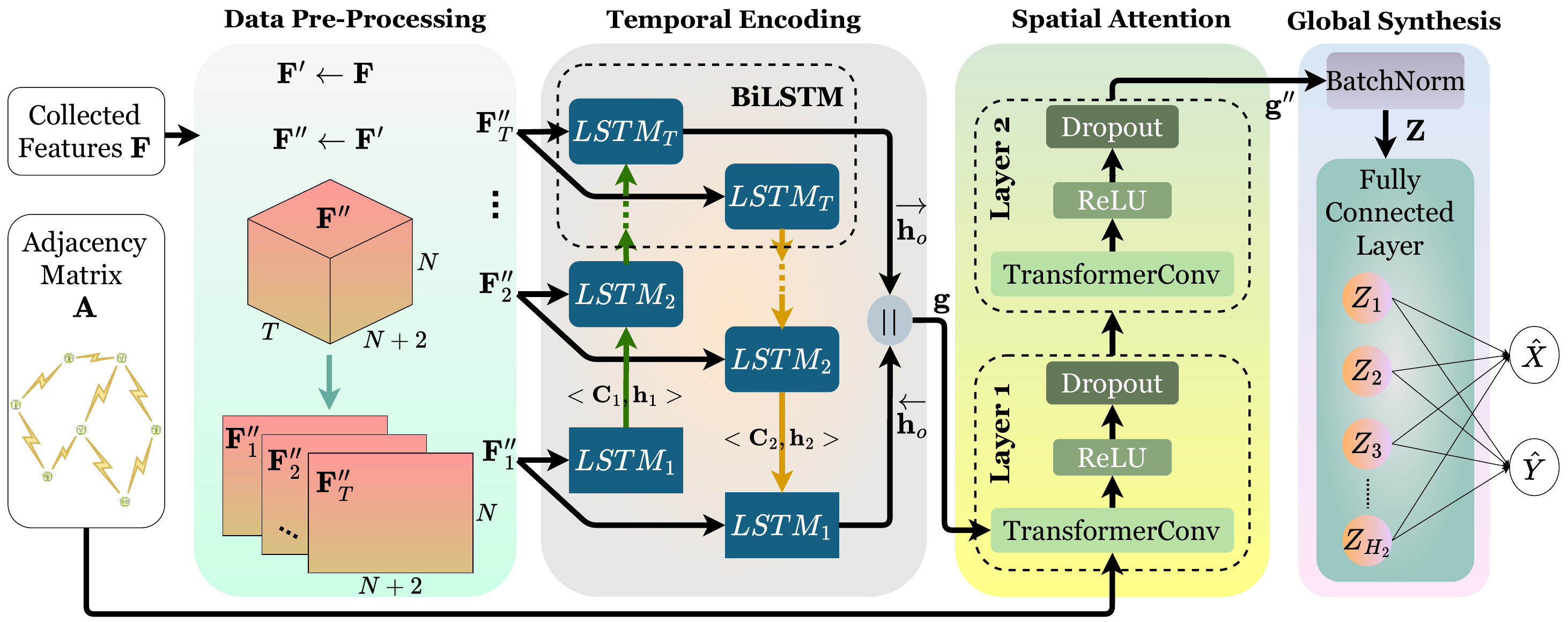}} 
      \caption{The proposed unified BiLSTM-Graph Transformer Localization framework for wireless IoT sensor networks, namely UBiGTLoc.}
      \label{model}
\end{figure*}

\section{UBiGTLoc: A Unified BiLSTM-Graph Transformer Localization  Framework for WSNs} \label{sec:UBiGTLoc}
We propose UBiGTLoc, a unified BiLSTM-Graph Transformer localization framework for sensor nodes in heterogeneous WSNs. UBiGTLoc is a unified approach that addresses the sensor nodes' localization problem under both anchor-presence and anchor-free scenarios. This allows to accommodate various IoT applications ranging from those of smart cites to smart agriculture in rural areas. 
UBiGTLoc is executed by a central node, i.e.,  a sink node, which promotes the  offloading of  computational extensive processes to the edge.

As shown in Fig.~\ref{model}, the proposed UBiGTLoc framework  comprises four main modules. These are the \textit{Data Pre-Processing Module}, the \textit{Temporal Encoding Module}, the \textit{Spatial Attention Module}, and the \textit{Global Synthesis Module}. In the \textit{Data Pre-Processing Module}, the central unit structures the feature matrix $\mathbf{F}$. This module ensures that the collected RSSI data is formatted properly, thereby setting the foundation for the subsequent processing.
The \textit{Temporal Encoding Module} employs a BiLSTM layer, which captures and summarizes the temporal dependencies inherent in the sequential RSSI measurements. 
The summarized output from the \textit{Temporal Encoding Module} along with the adjacency matrix \(\mathbf{A}\) is then fed into the third module of the proposed UBiGTLoc framework, i.e., the \textit{Spatial Attention Module}.  The \textit{Spatial Attention Module} consists of two Graph Transformer layers  to learn spatial patterns by  leveraging the output from the \textit{Temporal Encoding Module} and  the adjacency matrix. While the first layer helps to understand the WSNs's structure, the second layer leverages information
from border sensor nodes. 
Following, the output representations are normalized using batch normalization to stabilize training and ensure consistent feature scaling in the \textit{Global Synthesis Module}. The normalized data is passed through a fully connected layer, which projects the high-dimensional learned features into a two-dimensional space, resulting in the predicted sensor node locations ($\hat{X}, \hat{Y}$). We next discuss the modules of the proposed UBiGTLoc framework in detail.

\begin{algorithm}
\caption{Data Pre-Processing} \label{preproc}
\begin{algorithmic}[1]
\STATE \textbf{Input:} \( \mathbf{F} \in \mathbb{R}^{N \times (N+2) \times T} \)
\STATE \textbf{Output:} \( \mathbf{F}''_t \) for \( t \in \{1, \dots, T\} \)

\STATE \textbf{Step 1: Mean Imputation}
\FOR{ \( i = \{1, \dots, N\} \) }
    \FOR{ \( k = \{1, \dots, N+2\} \) }
        \FOR{ \( t = \{1, \dots, T\} \) }
            \IF{\( \text{f}_{i,k,t} \) is missing}
                \STATE \( \text{f}'_{i,k,t} \leftarrow \frac{1}{\tau} \sum_{t=1}^{\tau} \text{f}_{i,k,t} \)
            \ELSE
                \STATE \( \text{f}'_{i,k,t} \leftarrow \text{f}_{i,k,t} \)
            \ENDIF
            
     \STATE   \( \mathbf{F}'(i,k,t) \leftarrow \text{f}'_{i,k,t} \)
        \ENDFOR
    \ENDFOR
\ENDFOR

\STATE \textbf{Step 2: Z-score Normalization}
\FOR{ \( k = \{1, \dots, N+2\} \) }
    \STATE \( \mu_k \leftarrow \frac{1}{N \times T} \sum_{i=1}^{N} \sum_{t=1}^{T} \text{f}'_{i,k,t} \)
    \STATE \( \sigma_k \leftarrow \sqrt{\frac{1}{N \times T} \sum_{i=1}^{N} \sum_{t=1}^{T} \left(\text{f}'_{i,k,t} - \mu_k\right)^2} \)
    \FOR{ \( i = \{1, \dots, N\} \) }
        \FOR{ \( t = \{1, \dots, T\} \) }
            \STATE \( \text{f}''_{i,k,t} \leftarrow \frac{\text{f}'_{i,k,t} - \mu_k}{\sigma_k} \)
            \STATE \( \mathbf{F}''(i,k,t) \leftarrow \text{f}''_{i,k,t} \)
        \ENDFOR
    \ENDFOR
\ENDFOR

\STATE \textbf{Step 3: 2D Features' Extraction}
\FOR{ \( t = \{1, \dots, T\} \) }
    \STATE \( \mathbf{F}''_t  \leftarrow \mathbf{F}''_{:, :, t} \)
\ENDFOR

\STATE \textbf{Return} \( \mathbf{F}''_t~\forall~t \in \{1, \dots, T\} \)
\end{algorithmic}
\end{algorithm}

\subsection{Data Pre-Processing Module}
Algorithm~\ref{preproc} outlines the data pre-processing steps applied to the collected features over time period $T$ from all sensor nodes, i.e., \( \mathbf{F} \in \mathbb{R}^{N \times (N+2) \times T} \). This module is crucial to ensure the quality and consistency of the input data, which directly impacts the performance of the proposed UBiGTLoc model. 
The data pre-processing consists of three steps as follows.
\subsubsection{Step 1: Mean Imputation}
In the first step, missing values in \( \mathbf{F} \) are imputed using mean imputation~\cite{imputation}.  Each missing value is replaced by the mean of the available values for that feature across $T$ time steps for \emph{i-th} sensor node. Hence, 
The value of feature~\( k \) for the \emph{i-th} sensor node at time step~\( t \) after the imputation process (i.e., \( \text{f}^\prime_{i,k,t} \)) can be represented as:
\begin{equation} \label{tau}
\text{f}^\prime_{i,k,t} = 
\begin{cases}
\text{f}_{i,k,t} & \text{if } \text{f}_{i,k,t} \text{ is not missing,} \\
\frac{1}{\tau} \sum_{t=1}^{\tau} \text{f}_{i,k,t} & \text{if } \text{f}_{i,k,t} \text{ is missing.}
\end{cases}
\end{equation}
\noindent In Eq.~\eqref{tau}, \(\tau\) represents the number of non-missing values for feature~\( k \) across the $T$ time steps for the \emph{i-th} sensor node. 

Mean imputation is chosen due to its simplicity, effectiveness, and computational efficiency in maintaining the statistical properties of the dataset. Unlike more complex methods, such as K-nearest neighbors 
and various other machine learning-based imputation techniques, mean imputation requires significantly less computational resources and time~\cite{imputation}. It effectively handles the randomness of missing data without introducing significant biases, making it particularly suitable for localization tasks in WSNs. 
In this way, each of these imputed values (i.e., \(\text{f}^\prime_{i,k,t}\)) updates the corresponding entry in \(\mathbf{F}\), resulting in the imputed matrix \( \mathbf{F'} \), where every value \( \mathbf{F'}(i,k,t) \) is present, as indicated in line 12 of Algorithm~\ref{preproc}.

\subsubsection{Step 2: Z-score Normalization}
In the second step, the features in \( \mathbf{F}^{\prime} \) are normalized using Z-score normalization~\cite{z-norm}. This normalization is performed by first calculating the mean $\mu_k$ and standard deviation $\sigma_k$ for each feature~\( k \) across all sensor nodes and all time steps as follows: 
\begin{equation}
\mu_k = \frac{1}{N \times T} \sum_{i=1}^{N} \sum_{t=1}^{T} \text{f}_{i,k,t}^{\prime}, 
\end{equation}
\begin{equation}
\quad \sigma_k = \sqrt{\frac{1}{N \times T} \sum_{i=1}^{N} \sum_{t=1}^{T} \left( \text{f}_{i,k,t}^{\prime} - \mu_k \right)^2}
\end{equation}
\noindent Using these values, each element in \( \mathbf{F}^{\prime} \) is then standardized in the following way: 
\begin{equation}
\text{f}_{i,k,t}^{\prime\prime} = \frac{\text{f}_{i,k,t}^{\prime} - \mu_k}{\sigma_k}.
\end{equation}
\noindent Z-score normalization ensures that each feature~\( k \) has a mean of zero and a standard deviation of one. This method is preferred over other methods, such as min-max scaling and log-scaling normalization, because it facilitates faster convergence and stabilization during training~\cite{normalization}. In this way, each of these normalized values (i.e., \(\text{f}_{i,k,t}^{\prime\prime}\))  updates the corresponding entries of the normalized matrix \( \mathbf{F''} \), where every value \(\mathbf{F}^{\prime\prime}(i,k,t) \) is present, as indicated in line 23 of Algorithm~\ref{preproc}.
\subsubsection{Step 3: 2D Features' Extraction} 
After processing the 3D features matrix \( \mathbf{F}^{\prime\prime} \), we need to represent the data in a format suitable for input to the second module of the proposed UBiGTLoc framework, i.e., the \textit{Temporal Encoding Module}. In the machine learning context, \( \mathbf{F}^{\prime\prime} \) is treated as a tensor, which is a multi-dimensional array representing features data. To prepare the data for temporal encoding, the 3D features tensor \( \mathbf{F}^{\prime\prime} \) is divided into temporal tensors. Therefore, we extract 2D feature temporal tensors \( \mathbf{F}''_t \) for each time step \( t \) spanning from \( t = 1 \) to \( t = T \). Each 2D tensor \( \mathbf{F}''_t \) represents the feature vectors of all sensor nodes at the specific time stamp \( t \) and can be obtained directly from the 3D tensor \( \mathbf{F}'' \) as follows:
\begin{equation}
\mathbf{F}''_t = \mathbf{F}''_{:,:,t},
\end{equation}
\noindent where \( \mathbf{F}''_t \in \mathbb{R}^{N \times (N+2)} \). 
As a result, the 2D tensor \( \mathbf{F}''_t \) for a specific time stamp \( t \) can be represented as follow:
\begin{equation}
\mathbf{F}''_t =
\begin{bmatrix}
\text{f}''_{1,1,t} & \text{f}''_{1,2,t} & \cdots & \text{f}''_{1,(N+2),t} \\
\text{f}''_{2,1,t} & \text{f}''_{2,2,t} & \cdots & \text{f}''_{2,(N+2),t} \\
\vdots & \vdots & \ddots & \vdots \\
\text{f}''_{N,1,t} & \text{f}''_{N,2,t} & \cdots & \text{f}''_{N,(N+2),t}
\end{bmatrix},
\end{equation}
\noindent where each row represents a sensor node, and each column contains the pre-processed feature values for that sensor node at time step~\( t \). 
We next discuss the \textit{Temporal Encoding Module}.

\subsection{Temporal Encoding Module}
As mentioned previously, RSSI values can fluctuate significantly due to various factors such as interference, obstacles, and environmental changes. Accurately capturing and modeling these temporal variations is essential for reliable localization, and this is where the BiLSTM  network excels. BiLSTM extends the capabilities of the standard LSTM by processing the input sequence in both forward and backward directions~\cite{BiLSTM}. 
Thus,  BiLSTM provides a comprehensive understanding of the temporal~context. As depicted in Fig.~\ref{model}, each input \( \mathbf{F}''_t \) is processed by a BiLSTM, which consists of two LSTM cells per time step: one for processing the sequence in the forward direction (from past to future) and the other for processing in the backward direction (from future to past). Therefore, for each input sequence of length \( T \), the temporal encoding module utilizes a total of \( 2T \) LSTM cells, with the factor of 2 arising from the need to handle both~directions. 

\begin{figure}[t!]
\centerline{\includegraphics[width = 0.5 \textwidth]{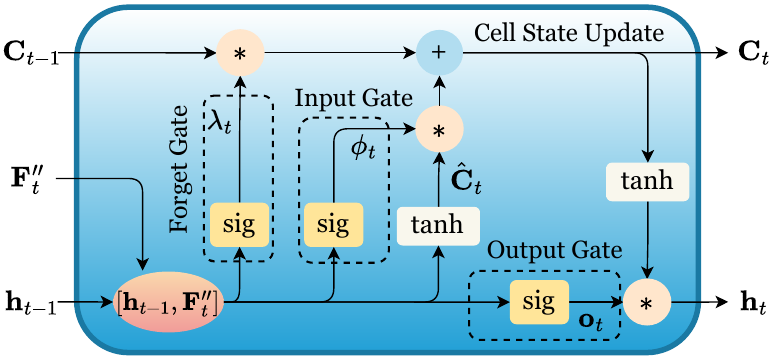}} 
      \caption{Computational graph of a single LSTM cell, i.e., $LSTM_t$.}
      \label{LSTM}
\end{figure} 

Fig.~\ref{LSTM} illustrates how each LSTM cell functions.  Each cell inputs are the pre-processed data at time stamp $t$ (i.e., \( \mathbf{F}''_t \)) and
the outputs from the previous time instance, which are the hidden state \(\mathbf{h}_{t-1}\) and the cell state \(\mathbf{C}_{t-1}\).  Inside each LSTM cell, a series of gating mechanisms  regulate the flow of information. These gates are the forget gate, the input gate, and the output gate. They allow the LSTM cell to maintain, update, or discard information as it propagates through the sequence to achieve the final outputs of cell state \(\mathbf{C}_{t}\) and hidden state \(\mathbf{h}_{t}\) as follows. 
\begin{enumerate}
    \item \textbf{Forget Gate}: At each time step \( t \), the forget gate decides which parts of the previous cell's state, denoted as \(\mathbf{C}_{t-1}\), should be retained. This decision is based on the previous hidden state \(\mathbf{h}_{t-1}\) from the previous time step (i.e., \( t-1 \)) and the current input \(\mathbf{F}_t''\). The forget gate output \(\mathbf{\lambda}_t\) is computed using the following equation: 
          \begin{equation}
            \mathbf{\lambda}_t = \text{sig} \left( \mathbf{W}_{\lambda} \times \left[ \mathbf{h}_{t-1}, \mathbf{F}_t'' \right] + \mathbf{\vec{b}}_{\lambda} \right),
          \end{equation}  
        \noindent where \(\mathbf{W}_{\lambda}\) is the weight matrix, and \(\mathbf{\vec{b}}_{\lambda}\) is the bias vector associated with the forget gate. 
        The sigmoid function \(\text{sig}(\cdot)\) ensures that \(\mathbf{\lambda}_t\) values range between 0 and 1, indicating the proportion of information to retain from the previous state \(\mathbf{C}_{t-1}\). Here, \(\times\) denotes matrix multiplication, and \(\left[\mathbf{h}_{t-1}, \mathbf{F}_t''\right]\) represents vertical concatenation. It is worth noting that for the very first cell (i.e., $LSTM_1$), \(\mathbf{C}_{t-1}\) and \(\mathbf{h}_{t-1}\) are initialized to~zeros.
    \item \textbf{Input Gate}: The input gate controls how much of the new information, represented by the candidate cell state \( \mathbf{\hat{C}}_t \), should be added to the current cell state \( \mathbf{C}_t \). 
    The candidate cell state \( \mathbf{\hat{C}}_t \) is generated using a tanh$(\cdot)$ activation function, which scales the new input information (i.e., \(\left[ \mathbf{h}_{t-1}, \mathbf{F}_t'' \right]\)) to a range suitable for combination with the cell state as follows:
    \begin{equation}
    \hat{\mathbf{C}}_t = \tanh \left( \mathbf{W}_C \times \left[ \mathbf{h}_{t-1}, \mathbf{F}_t'' \right] + \mathbf{\vec{b}}_C \right),
    \end{equation}
    \noindent where \(\mathbf{W}_C\) is the weight matrix, and \(\mathbf{\vec{b}}_C\) is the bias vector associated with the candidate cell state computation.
    Meanwhile, the input gate processes the current input \(\mathbf{F}_t''\) and the previous hidden state \(\mathbf{h}_{t-1}\) through a sigmoid activation function to produce:
    \begin{equation}
    \mathbf{\phi}_t = \text{sig} \left( \mathbf{W}_{\phi} \times \left[ \mathbf{h}_{t-1}, \mathbf{F}_t'' \right] + \mathbf{\vec{b}}_{\phi} \right),
    \end{equation}
    where \(\mathbf{\phi}_t\) is the input gate output at time step \(t\), \(\mathbf{W}_{\phi}\) is the weight matrix, and \(\mathbf{\vec{b}}_{\phi}\) is the bias vector associated with the input gate.
    \item \textbf{Cell State Update}: The cell state \( \mathbf{C}_t \) is updated by combining the retained information from the previous state \( \mathbf{C}_{t-1} \) with the new information from the candidate cell state \( \mathbf{\hat{C}}_t \). The forget gate output \(\mathbf{\lambda}_t\) dictates the retention of \( \mathbf{C}_{t-1} \) , while the input gate  output \(\mathbf{\phi}_t\) governs the addition of \( \mathbf{\hat{C}}_t \).  The cell state \( \mathbf{C}_t \) is then calculated as follows:
    \begin{equation}
    \mathbf{C}_t = \lambda_t * \mathbf{C}_{t-1} + \mathbf{\phi}_t * \hat{\mathbf{C}}_t,
    \end{equation}
    \noindent where \(*\) denotes element-wise multiplication. 
    \item \textbf{Output Gate}: The output gate determines which part of the cell state \(\mathbf{C}_t\) will be output as the new hidden state \(\mathbf{h}_t\) and will be used in the next time step or fed into subsequent layers of  the network. The output gate output $\mathbf{o}_t$ is computed in the following way:
    \begin{equation}
    \mathbf{o}_t = \text{sig}(\mathbf{W}_o \times [\mathbf{h}_{t-1}, \mathbf{F}''_t] + \mathbf{\vec{b}}_o),
    \end{equation}
    where 
    \(\mathbf{W}_o\) is the weight matrix, and \(\mathbf{\vec{b}}_o\) is the bias vector associated with the output gate. 
    The hidden state \(\mathbf{h}_t\) is then produced by applying a tanh$(\cdot)$ activation function to the updated cell state, scaled by \(\mathbf{o}_t\) as follows:
    \begin{equation}
    \mathbf{h}_t = \mathbf{o}_t * \tanh(\mathbf{C}_t).
    \end{equation}
\end{enumerate}

As shown in Fig.~\ref{model}, from the sequence of hidden states generated by the BiLSTM, we select the hidden states of the last LSTM cell 
in each direction of the BiLSTM. These are denoted as \(\overrightarrow{\mathbf{h}}_o\) for the forward direction and \(\overleftarrow{\mathbf{h}}_o\) for the backward direction. Both \(\overrightarrow{\mathbf{h}}_o\) and \(\overleftarrow{\mathbf{h}}_o\) summarize the information from their respective directions.
In contrast to conventional methods that utilize the output of all hidden states from each time step, this article proposes using only the final hidden states from each direction. This approach provides a holistic summary of the entire sequence by capturing information from both the forward and backward passes. By selecting the final hidden states, we effectively consolidate all temporal information into a compact representation, which helps in focusing on the overall sequence context. This method reduces dimensionality and computational complexity while retaining the comprehensive context of the input sequence.
As shown in Fig.~\ref{model}, the summarized hidden states from both directions are then concatenated to form a comprehensive temporal encoding, denoted as \(\mathbf{g}\). 
The equation for \(\mathbf{g}\) is given~by:
\begin{equation}
\mathbf{g} = [\overrightarrow{\mathbf{h}}_o \, || \, \overleftarrow{\mathbf{h}}_o] = \left[\mathbf{\vec{g}}_1; \mathbf{\vec{g}}_2; \dots; \mathbf{\vec{g}}_N\right]^T,
\end{equation}
\noindent where \(||\) denotes horizontal concatenation, and \(\mathbf{\vec{g}}_i\) represents the row vector of feature outputs for the \textit{i-th} sensor node. 
Considering \( H_1 \) as a hidden dimension of each LSTM cell, which represents the number of output features in the hidden state \(\mathbf{h}_t\), the tensor \(\mathbf{g}\) has a dimensionality of \(\mathbb{R}^{N \times 2H_1}\), encapsulating the summarized information from all time steps. 
The tensor~\(\mathbf{g}\) serves as the input for the subsequent \textit{Spatial Attention Module}, providing a powerful representation that captures both short-term and long-term dependencies. Following, we discuss the \textit{Spatial Attention~Module}.


\begin{figure}[t!]
\centerline{\includegraphics[width = 0.43 \textwidth]{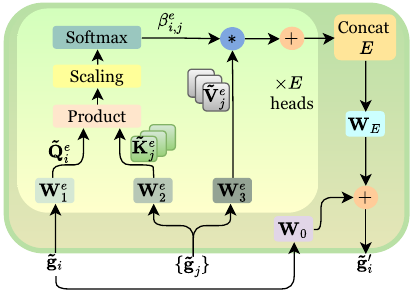}} 
      \caption{The computational graph of TransformerConv.}
      \label{TransformerConv}
\end{figure} 

\subsection{Spatial Attention Module}
As shown in Fig.~\ref{model}, the \textit{Spatial Attention Module}  processes the sensor nodes' features obtained from the previous module (i.e., \(\mathbf{g}\)) and the adjacency matrix \(\mathbf{A}\) to effectively learn the spatial relationships between sensor nodes. The \textit{Spatial Attention Module} starts with a TransformerConv operation of Layer~1. TransformerConv is a GNN variant of the transformer architecture designed for processing graph-structured data~\cite{shi_masked_2021}. Fig.~\ref{TransformerConv} shows the  computational graph of a TransformerConv process. A TransformerConv operation combines sensor nodes' feature aggregation with multi-head attention, which allows the model to focus on the most relevant neighbors' information effectively. Multi-head attention employs~$E$ heads to capture different aspects of sensor node relationships, which improves feature aggregation by considering multiple perspectives simultaneously. This is done by computing the attention-weighted sum of the neighboring sensor node features. 
For this purpose, the attention coefficient $\beta^e_{i,j}$ for each head $e$ is computed as part of the multi-head attention mechanism. In this process, each feature vector $\mathbf{\vec{g}}_i \in \mathbb{R}^{H_1}$ of the \emph{i-th} sensor node is transformed into a query vector $\mathbf{\vec{Q}}_i^e \in \mathbb{R}^{H_2}$, which is given by:
\begin{equation}
   \mathbf{\vec{Q}}_i^e = \mathbf{W}_1^e \times \mathbf{\vec{g}}_i, 
\end{equation}
\noindent where $\mathbf{W}_1^e \in \mathbb{R}^{H_2 \times H_1}$ is the weight matrix for each attention head $e$, with $H_2$ being the hidden dimension of the TransformerConv layer.
Similarly, the feature vector $\mathbf{\vec{g}}_j \in \mathbb{R}^{H_1}$ of each neighboring sensor node $j$ is transformed into a key vector $\mathbf{\vec{K}}_j^e \in \mathbb{R}^{H_2}$ using another weight matrix $\mathbf{W}_2^e \in \mathbb{R}^{H_2 \times H_1}$, specific to the \emph{e-th} head in the following way:
\begin{equation}
    \mathbf{\vec{K}}_j^e = \mathbf{W}_2^e \times \mathbf{\vec{g}}_j.
\end{equation}
The attention score between sensor node $i$ and each of its neighbors $j$ for head $e$ is calculated by taking the dot product of the query $\mathbf{\vec{Q}}_i^e$ and key $\mathbf{\vec{K}}_j^e$, scaled by the square root of their dimensionality $H_2$. This score is then normalized across all neighbors using a softmax function to produce the attention coefficient $\beta^e_{i,j}$ as follows:
\begin{equation}
\beta^e_{i,j} = \text{softmax} \left( \frac{(\mathbf{\vec{Q}}_i^e)^{\top} \mathbf{\vec{K}}_j^e}{\sqrt{H_2}} \right).
\end{equation}
The value vector $\mathbf{\vec{V}}_j^e \in \mathbb{R}^{H_2}$ is obtained by passing the feature vector of the neighboring sensor node $j$ through a weight matrix $\mathbf{W}_3^e \in \mathbb{R}^{H_2 \times H_1}$ for each attention head $e$ as follows:
\begin{equation}
    \mathbf{\vec{V}}_j^e = \mathbf{W}_3^e \times \mathbf{\vec{g}}_j.
\end{equation}
The value vectors are then weighted by their corresponding attention coefficients $\beta^e_{i,j}$ and summed for each attention head. The outputs from all $E$ attention heads are then concatenated (i.e., $\bigoplus$) to form a single vector, which is then passed through the linear projection matrix $\mathbf{W}_{E} \in \mathbb{R}^{H_2 \times (E H_2)}$ to reduce its dimensionality back to the original size.
This projection step ensures that the updated feature vector for the target sensor node retains the same output dimensionality \(H_2\) of one head, while incorporating information from multiple attention heads. The output is then used to update the target sensor node's feature $\mathbf{\vec{g}}^{\prime}_i \in \mathbb{R}^{H_2}$ as follows:
\begin{equation}
\mathbf{\vec{g}}^{\prime}_i = \mathbf{W}_0 \times \mathbf{\vec{g}}_i + \mathbf{W}_{E} \times \left[ \bigoplus_{e=1}^{E} \left( \sum_{j \in \mathcal{N}(i)} \beta^e_{i,j} \mathbf{\vec{V}}_j^e \right) \right],
\end{equation}
\noindent where the feature vector of a sensor node $\mathbf{\vec{g}}_i$ is passed through a learnable weight matrix $\mathbf{W}_0 \in \mathbb{R}^{H_2 \times H_1}$ and combined with the output of the multi-head attention mechanism.


This process is applied simultaneously for all sensor nodes in the network, producing a combined tensor \( \mathbf{g}' \in \mathbb{R}^{N \times H_2} \), where each row represents the updated feature vector \( \mathbf{\vec{g}}^{\prime}_i \) of an individual sensor node. As illustrated in Fig.~\ref{model}, this tensor \( \mathbf{g}' \) is subsequently passed through a Rectified Linear Unit (ReLU) activation function in Layer 1. The ReLU activation introduces non-linearity into the model, which is crucial for capturing more complex spatial patterns. To further enhance the robustness of the model and mitigate the risk of overfitting, a dropout layer is applied after the ReLU activation. This dropout layer randomly zeroes out a fraction of units during training, effectively preventing the model from becoming too reliant on specific~features.

Next, the output from Layer~1 is passed into Layer~2. Layer~2 is identical in structure to Layer 1. It helps the model in refining spatial relationships, to leverage information from border sensor nodes. 
Thus, the final transformed features, denoted as \( \mathbf{g}^{\prime\prime} \in \mathbb{R}^{N \times H_2} \), represent the learned spatial information for all sensor nodes. These final features are then passed to the \textit{Global Synthesis Module}, which is discussed in detail in the following section.

\subsection{Global Synthesis Module} \label{sub:global_synthesis}
The \textit{Global Synthesis Module} integrates the outputs from the \textit{Spatial Attention Module} to produce the final predicted coordinates of sensor nodes. This module ensures that the high-dimensional features learned by the previous layers are synthesized into meaningful spatial coordinates, representing the precise location of each sensor node. It comprises of two steps, i.e., batch normalization and a fully connected layer.
\subsubsection{Batch Normalization}
The first step in the \textit{Global Synthesis Module} involves applying Batch Normalization to the output representations from the \textit{Spatial Attention Module}, i.e.,~\( \mathbf{g}'' \). Batch normalization is critical for stabilizing the training process by normalizing the feature distributions across each mini-batch of training examples \cite{normalization}. Here, a batch refers to a subset of the dataset processed simultaneously by the model during training. 
Using Batch Normalization, the input is first scaled as follows:
\begin{equation}
    \hat{\mathbf{g}}^{\prime\prime} = \frac{\mathbf{g}^{\prime\prime} - \mu_B}{\sqrt{\sigma_B^2 + \epsilon}},
    \end{equation}
\noindent where \( \mu_B \) and \( \sigma_B^2 \) are the mean and variance of the current mini-batch, respectively, and \( \epsilon \) is a small constant added to avoid division by zero.
    The normalized output \( \hat{\mathbf{g}}^{\prime\prime} \in \mathbb{R}^{N \times H_2} \) is then shifted to produce the final output \( \mathbf{Z} \in \mathbb{R}^{N \times H_2} \) in the following way:
    \begin{equation}
    \mathbf{Z} = \gamma * \hat{\mathbf{g}}^{\prime\prime} + \delta.
\end{equation}
\noindent The learnable parameters \( \gamma \) and \( \delta \) allow the model to scale and shift the normalized output, restoring the representation’s expressive power after normalization.
By applying batch normalization after the \textit{Spatial Attention Module}, the model ensures that the transformed features maintain consistent statistical properties across batches. This consistency is crucial for effective learning in the subsequent fully connected layer, as it prevents the features from shifting their distribution during training. Thus, batch normalization helps in mitigating the internal covariate shift, leading to a more stable and efficient training process.
\subsubsection{Fully Connected Layer}
After stabilizing the feature distributions with batch normalization, the \textit{Global Synthesis Module} projects these normalized features into the final two-dimensional space using a fully connected layer. The purpose of this layer is to synthesize the high-dimensional learned features into \( \hat{X} \) and \( \hat{Y} \), which represent the predicted locations of the sensor nodes.
The fully connected layer is mathematically expressed as:
\begin{align}
    \mathbf{\left(\hat{X} ,\hat{Y}\right)} = \mathbf{W} \times \mathbf{Z} + \mathbf{\vec{b}}
\end{align}
\noindent where \( W \) is the weight matrix and \( b \) is the bias vector. As depicted in Fig.~\ref{model}, the fully connected layer has an input dimension of \(H_2\), which corresponds to the dimensionality of the feature vector output from the batch normalization layer i.e., \( \mathbf{Z} \). The output dimension of this layer is  $N\times2$, representing the predicted x and y coordinates for the sensor nodes, i.e., \( \hat{X} \) and \( \hat{Y} \).

The design of the \textit{Global Synthesis Module}, which combines normalization with direct projection into the coordinate space, plays a pivotal role in the overall accuracy and robustness of the WSN localization model. This module acts as the final module in our architecture, ensuring that the learned spatial dependencies are effectively translated into real-world coordinates, suitable for deployment in various IoT scenarios.


\section{Simulation Settings and Training Process} \label{sec:sim_settings}
This section outlines the simulation and training configurations used to evaluate the performance of the proposed UBiGTLoc approach. 
The training procedure, including data preparation, model hyperparameters, and optimization strategies, are also detailed.

\subsection{Simulation Settings}
We conduct our simulations using Python 3.11.8~\cite{python}, with the key parameters detailed in Table~\ref{sim_train_settings}. The simulation area covers \(100 \times 100~\text{m}^2\), with sensor nodes randomly distributed within the WSN field. The number of sensor nodes \(N\) varies from \(100\) to \(500\). To evaluate the impact of anchor nodes, we experiment various anchor node percentages in WSNs, i.e., \(\alpha\) ranging from \(0\%\) (anchor-free) to \(50\%\). We also examine the effect of radio range on sensor nodes' localization by varying the radio range distance threshold \(d_{\text{th}}\) between \(2\) and \(100\) meters. We consider the time window size \(T\) from \(3\) to \(30\) timestamps to identify the optimal duration. Gaussian noise with zero mean and variance \(\sigma^2\) ranging from \(0.04\) to \(0.5\) is used to simulate RSSI fluctuations. 

In most of our simulations, we use fixed parameters unless we are specifically studying the effect of changing a particular parameter. We consider two scenarios, a sparse network where \(N = 100\) and a dense network where \(N = 500\).
The anchor node ratio \(\alpha\) is fixed at \(20\%\), the radio range distance threshold \(d_{\text{th}}\) is set to \(20\) meters, the time window size \(T\) is set to \(10\) timestamps, the Gaussian noise variance \(\sigma^2\) is set to \(0.5\). The interference scaling factor $\kappa$, which takes the value between zero and one is set to zero to isolate the effects of density-dependent interference, unless specified~otherwise.

\begin{table}[t!]
\centering
\caption{Simulations and Training Settings}
\label{sim_train_settings}
\renewcommand{\arraystretch}{1.25}
\begin{tabular}{|l|l|}
\hline
\textbf{Parameter}                & \textbf{Value}            \\ \hline
\(L \times L\) (WSN field area)             & \(100 \times 100~\text{m}^2\) \\ \hline
\(N\) (Number of sensor nodes)           & \(100 - 500\)             \\ \hline
\(\alpha\) (The percentage of anchor  nodes)         & \(0 - 50 ~\%\)            \\ \hline
\(d_{th}\) (Radio range)          & \(2 - 100\) m             \\ \hline
\(T\) (Time window)               & \(3 - 30\)                \\ \hline
\(\sigma^2\) (RSSI noise)         & \(0.04 - 0.5\)        \\ \hline
\(\kappa\) (Interference scaling factor)         & \(0 - 1\)       \\ \hline
Number of WSN graph samples                 & \(1000\)           \\ \hline
Training / testing size          & 80\% / 20\%                      \\ \hline
K-fold cross-validation           & 5                         \\ \hline
Data augmentation             & 50\%                      \\ \hline
Edge removal   percentage                   & 10\%                      \\ \hline
Noise addition         & 0.1            \\ \hline
Dropout   percentage                          & 50\%            
\\ \hline
Batch size                        & 16                        \\ \hline
Number of epochs                            & 100                       \\ \hline
Learning rate                     & 0.001                     \\ \hline
\end{tabular}
\end{table}

\subsection{Training Settings and Procedure}
For the training and evaluation of our model, we utilize PyTorch Geometric (PyG)~2.2.2 \cite{fey2019pytorch}, a library built on PyTorch that facilitates 
training of GNNs. 
Table~\ref{sim_train_settings} provides a summary of the training hyperparameters used. These settings and hyperparameters are selected based on extensive experimentation to achieve the best balance between model performance, training efficiency, and generalization across various WSN topologies.

To begin with, our dataset consists of \(1000\)~graph samples, generated as described in Section~II. Specifically, we consider \(100\)~distinct WSN topologies, with each topology exhibiting different sensor node positions and connectivity patterns. For each of these topologies, we then generate \(10\)~different graph samples by sampling from Gaussian noise to simulate the RSSI variations. This setup aims to replicate real-world conditions, including scenarios such as sensor node mobility, sensor node failures, and varying connectivity. To evaluate the model effectively and prevent overfitting, we first split the dataset into $80\%$ for training and $20\%$ for testing, ensuring that the testing set remains independent throughout the model development process. The training data is then further subjected to 5-fold cross-validation~\cite{cross-validation}. In this process, the training set is split into five equal parts, with the model being trained on four folds and validated on the remaining fold. This is repeated across all folds, and the results are averaged to assess overall performance. After the cross-validation process, the model is retrained on the entire training set using the optimal hyperparameters. Final evaluation is performed on the independent test set, ensuring generalization across different network configurations. During the training process, we further utilize data augmentation and dropout as well as the mini-batch gradient descent approach in the following way.
\subsubsection{Data Augmentation and Dropout}
We implement data augmentation to further enhance the model's robustness. Specifically, we apply two types of graph augmentations, \textit{structure-oriented} and \textit{feature-oriented}, with a 50\% probability for~each~\cite{Data-augmentation}:
\begin{itemize}
\item \textit{Structure Oriented (i.e., Edge Removal)}: To simulate varying levels of WSN  connectivity, we randomly remove $10\%$ of the edges in the graph samples of our dataset. This augmentation helps the model to better generalize  to scenarios where connectivity is incomplete or fluctuates.
\item \textit{Feature Oriented (i.e., Noise Addition)}: We add extra Gaussian noise with a mean of zero and a standard deviation of $0.1$ to the RSSI features. This simulates real-world measurement errors and helps the model adapt to noise and uncertainties in the data.
\end{itemize}
In addition to data augmentation, we apply dropout during training with a probability of $50\%$~\cite{srivastava2014dropout}. Dropout is a regularization technique that randomly deactivates a subset of neurons during training, thus, preventing the model from becoming overly reliant on specific features and thus reducing overfitting.
\subsubsection{Mini-Batch Gradient Descent}
The training procedure further utilizes the mini-batch gradient descent approach, which involves dividing the training dataset into smaller batches of samples~\cite{mini-batches}. This technique allows the model to update its parameters more frequently than full-batch training, leading to faster convergence and more stable updates. Empirically, we select a batch size of $16$, which strikes a balance between efficient use of computational resources and maintaining the stability of the model during training. For optimization, we adopt Adam optimizer, known for its effectiveness in training deep learning models~\cite{kingma_adam_2017}. The learning rate is set to $0.001$ to ensure steady progress toward minimizing the loss function. The end-to-end training is done over $100$~epochs, which provides an opportunity for the model to converge to an optimal solution.
\subsection{UBiGTLoc Hyperparameters Tuning}
As described in Section~III, apart from the \textit{Data Pre-Processing Module},  UBiGTLoc consists of three key modules: the \textit{Temporal Encoding Module}, the \textit{Spatial Attention Module}, and the \textit{Global Synthesis Module}. The hyperparameters for each module are empirically selected based on extensive experimentation. We summarize these choices below.
\subsubsection{Temporal Encoding Module Hyperparameters}
We found that a hidden size of \(H_1 = 500\)  with a single BiLSTM layer is sufficient to provide the necessary temporal information for accurate localization. This choice improves computational efficiency, thus, resulting in faster training and inference without compromising the~performance.
\subsubsection{Spatial Attention Module Hyperparameters}
For the TransformerConv process, four attention heads, i.e., \(E = 4\), and a hidden dimension of \(H_2 = 500\) were identified  to strike a balance between learning complex patterns and maintaining computational efficiency.
\subsubsection{Global Synthesis Module Hyperparameters}
The global synthesis module applies a batch normalization layer with an input size equal to the output dimension of the spatial attention module, i.e., \(H_2 = 500\). Following this, a fully connected layer maps the \(H_2 = 500\) features to a 2-dimensional output~\((\hat{X}, \hat{Y})\), which represent the predicted sensor nodes' locations.

Following, we discuss the simulation results.

\begin{table}[b!]
\caption{The average localization loss under different noise levels}
\label{comparison_baselines}
\centering
\renewcommand{\arraystretch}{1.1} 
\setlength{\tabcolsep}{3.5pt} %
\begin{tabular}{|l|c|c|c|c|c|c|c|c|c|}
\hline
\textbf{Methods} & \textbf{$N$} & \multicolumn{4}{c|}{\textbf{Noise ($\sigma^2$)}} \\ 
\cline{3-6}
& & \textbf{0.04} & \textbf{0.1} & \textbf{0.25} & \textbf{0.5} \\
\hline
\multirow{2}{*}{\textbf{GCN} \cite{yan_graph_2021}} 
& 500  & 5.5168 & 6.0015 & 6.1833 & 6.2045 \\
& 100  & 10.7632 & 11.1269 & 11.7125 & 12.2493 \\
\hline
\multirow{2}{*}{\textbf{U-MLP} \cite{ahmed_unified_2024}} 
& 500  & 3.6481 & 3.8112 & 3.9057 & 4.2253 \\
& 100  & 7.6187 & 8.1471 & 8.6731 & 9.0554 \\
\hline
\multirow{2}{*}{\textbf{AGNN} \cite{yan_attentional_2023}} 
& 500  & 1.6538 & 1.6819 & 1.7115 & 1.7416 \\
& 100  & 5.2864 & 5.7294 & 6.0452 & 6.4973 \\
\hline
\multirow{2}{*}{\textbf{Baseline 1}} 
& 500  & 1.3120 & 1.3410 & 1.3777 & 1.4101 \\
& 100  & 4.9437 & 5.2332 & 5.5674 & 5.8173 \\
\hline
\multirow{2}{*}{\textbf{Baseline 2}} 
& 500  & 1.0947 & 1.1378 & 1.1719 & 1.2158 \\
& 100  & 2.8624 & 3.0924 & 3.4714 & 3.7415 \\
\hline
\multirow{2}{*}{\textbf{UBiGTLoc \((\alpha = 0)\)}} 
& 500  & 0.4464 & 0.4783 & 0.5001 & 0.5324 \\
& 100  & 1.1635 & 1.3472 & 1.5271 & 1.7191 \\
\hline
\multirow{2}{*}{\textbf{UBiGTLoc \((\alpha = 20\%)\)}} 
& 500  & 0.3721 & 0.4013 & 0.4318 & 0.4740 \\
& 100  & 0.8247 & 0.9871 & 1.0741 & 1.2523 \\
\hline
\end{tabular}
\end{table}

\section{Simulation Results} \label{sec:sim_results}
In this section, we evaluate the proposed UBiGTLoc framework and compare its performance with existing methods: GCN \cite{yan_graph_2021}, U-MLP \cite{ahmed_unified_2024}, and AGNN \cite{yan_attentional_2023}, which rely on anchor nodes for localization (see Section~\ref{sec:related_work}). The GCN method uses a two-layer graph convolutional network with $2000$~hidden units, relying on a distance threshold to maintain graph topology and Gaussian noise for handling interference. U-MLP utilizes MLP without leveraging graph structure, while AGNN employs two GATv2 (Graph Attention Networks) layers with $2000$~hidden units and a learned adjacency module based on the attention mechanism.

We also define two baselines derived from UBiGTLoc for comparison and ablation studies, to demonstrate the significance of the temporal encoding module within UBiGTLoc:
\begin{enumerate}
    \item Baseline~1 excludes the temporal encoding module and processes only a single RSSI snapshot, specifically the last timestamp features, i.e., \( \mathbf{F}'_T \), as input to the spatial attention module. In this baseline, the spatial attention module consists of two layers with a hidden dimension \(H_2 = 500\) per layer.
    \item Baseline~2 employs the Exponential Weighted Moving Average (EWMA) within the temporal encoding module, replacing the BiLSTM. EWMA processes input features from multiple time steps by assigning exponentially decreasing weights to older observations, prioritizing more recent data \cite{EWMA}. As with Baseline 1, two subsequent layers are used in the spatial attention module. 
\end{enumerate}

The performance of all methods is evaluated and compared based on the MSE as the primary metric (discussed in Section~II-A). 
We further analyze how various factors influence localization performance. These factors include the distance threshold, sensor node density, time window size, and the percentage of anchor nodes available in the WSN. All the figures were plotted using Matlab~R2024b~\cite{matlab}. 

\subsection{RSSI Imperfections Variations Effect on MSE}
Table~\ref{comparison_baselines} presents the average localization loss (i.e., MSE) in meters under various RSSI noise levels with  \(\sigma^2\) ranging between  \(0.04\) and \(0.5\). The results compare our proposed UBiGTLoc model to existing methods (i.e., GCN, U-MLP, AGNN) and the two ablation baselines under two WSN sizes (i.e., 500 and 100 sensor nodes).

Among all methods, UBiGTLoc with anchors (i.e., \(\alpha=20\%\)) achieves the best performance with the error of \(0.4740\)~m in dense network settings and \(1.2523\)~m in sparse network settings under the worst noise scenario of \(\sigma^2 = 0.5\). This demonstrate its ability to handle noise while maintaining high localization accuracy. Even in the anchor-free scenario (i.e., \(\alpha=0\)), UBiGTLoc consistently outperforms  other methods,  showing lower error across all noise levels and comparable results with the anchor-presence scenario. It results in the error of  \(0.5324\)~m in dense network settings and \(1.7191\)~m in sparse network settings under the worst noise scenario of \(\sigma^2 = 0.5\). This  highlights UBiGTLoc's robustness in scenarios where anchor nodes are unavailable. The incorporation of BiLSTM allows UBiGTLoc to better capture temporal dependencies compared to the EWMA used in Baseline 2. Baseline 2 shows the error results of \(1.2158\)~m in dense network settings and \(3.7415\)~m in sparse network settings under the worst noise scenario of \(\sigma^2 = 0.5\). UBiGTLoc with and without anchor nodes as well as  Baseline 2 stand out because these methods incorporate temporal encoding and significantly outperform those that lack temporal encoding. Baseline 1 achieves slightly better results than AGNN by \(0.3
-0.6\)~m, yet its performance degrade in noisy environments due to the lack of temporal information. Although AGNN outperforms U-MLP and GCN, it  fails to match the performance of more advanced methods due to the absence of temporal encoding. U-MLP, lacking a graph structure, performs slightly better by \(2-3\)~m than GCN but still shows significant localization errors compared to graph-based models, particularly those with attention mechanisms. GCN 
exhibits the highest localization errors across all noise levels under both sparse and dense network settings.  
It is worth noting that as noise levels increase, all methods experience a degradation in performance, albeit to varying degrees. For instance, UBiGTLoc in both anchor-presence and anchor-free scenarios exhibits a modest increase in loss of only \(0.1\)~m when \(N = 500\) and \(0.5\)~m when \(N = 100\) as \(\sigma^2\) rises from \(0.04\) to \(0.5\). This indicates a higher immunity to noise compared to the existing methods.  In contrast, GCN experiences more substantial jumps in localization error of approximately \(0.7\)~m and \(1.5\)~m for \(N = 500\) and \(N = 100\), respectively.


\begin{figure}[t!]
    \centerline{\includegraphics[width=0.5\textwidth]{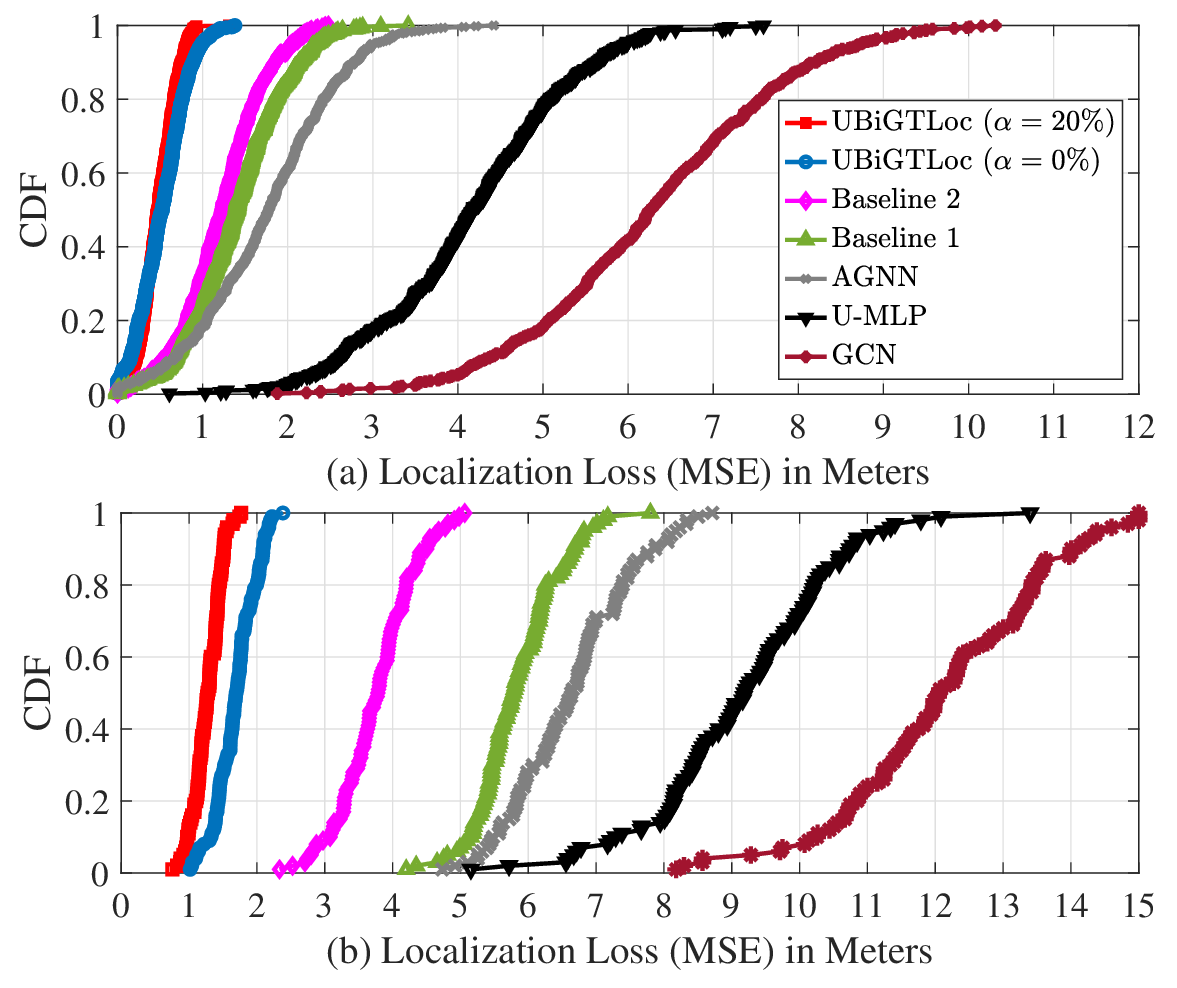}}
    \caption{CDF versus localization loss (in meters) for two network settings: (a) 500 sensor nodes, (b) 100 sensor nodes.}
    \label{Fig:CDF}
\end{figure}

To further investigate the performance under noisy environments, Fig.~\ref{Fig:CDF}(a) and  Fig.~\ref{Fig:CDF}(b) present the cumulative distribution function (CDF) of sensor nodes localization loss for all considered methods at a fixed noise level with \(\sigma^2 = 0.5\), evaluated over the network settings of 500 and 100 sensor nodes, respectively. The CDF plot illustrates the distribution of localization loss among sensor nodes in the WSN. For instance, in Fig.~\ref{Fig:CDF}(a), the UBiGTLoc (\(\alpha=20\%\)) shows that around 90\% of the sensor nodes achieve a localization loss of less than \(1\)~m, while its anchor-free counterpart (\(\alpha=0\)) still maintains comparable  accuracy, outperforming all other methods, including those relying on anchor nodes.
GCN exhibits the worst performance, with a much flatter curve with only 20\% of the sensor nodes achieve a localization loss of less than \(5\)~m, indicating a higher localization loss for most of the sensor nodes. While AGNN performs better than U-MLP and GCN with 80\% of the sensor nodes achieve a localization loss of less than \(2.5\)~m, it still cannot match the precision of UBiGTLoc in the anchor-free scenario. Baseline 1, achieves slightly better than AGNN, but it lags behind Baseline~2. Even though Baseline 2 improves accuracy, it remains inferior to the temporal processing capabilities of UBiGTLoc's BiLSTM-based approach. 
In Fig.~\ref{Fig:CDF}(b), which depicts an 100 sensor node WSN, the same trend is observed, but with generally higher localization losses across all methods. For UBiGTLoc (\(\alpha=20\%\)), around 80\% of the sensor nodes achieve a localization loss of less than \(1.5\)~m, while the anchor-free variant (\(\alpha=0\)) maintains comparable performance with around 80\% of sensor nodes achieving a loss below \(2\)~m. Other methods, such as GCN and U-MLP, show more significant degradation, with flatter curves, indicating a higher localization loss for most of the sensor nodes. 


\begin{figure}[t!]
    \centerline{\includegraphics[width=0.5\textwidth]{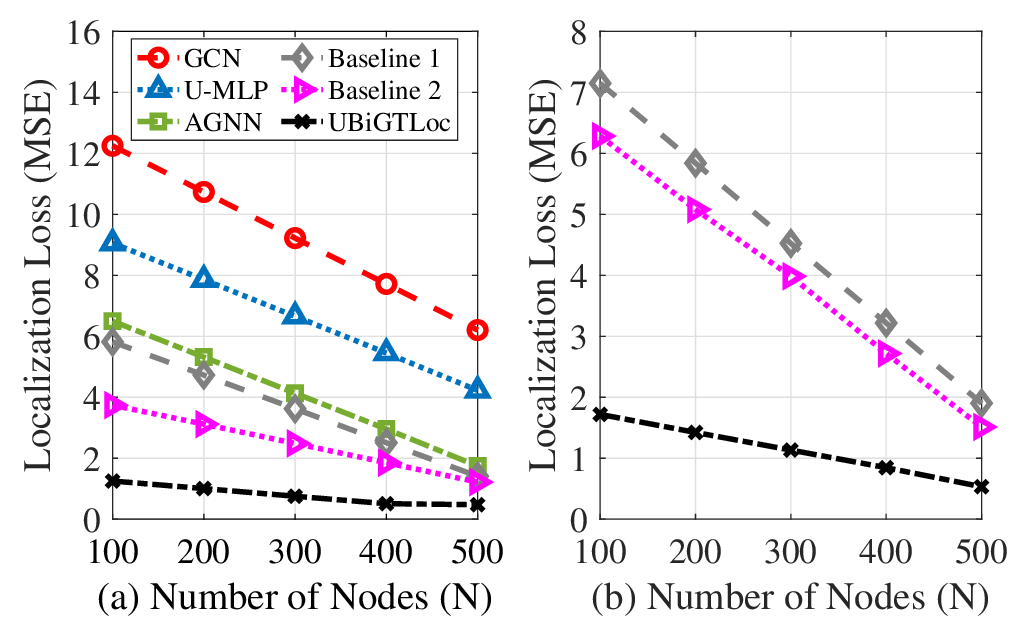}}
    \caption{Localization loss versus the number of sensor nodes in an ideal  interference-free scenario (i.e., $\kappa = 0$) 
    for two different scenarios: (a) anchor-presence ($\alpha = 20\%$), (b) anchor-free ($\alpha = 0$).}
    \label{fig:density effect}
\end{figure}

\begin{figure}[t!]
    \centerline{\includegraphics[width=0.5\textwidth]{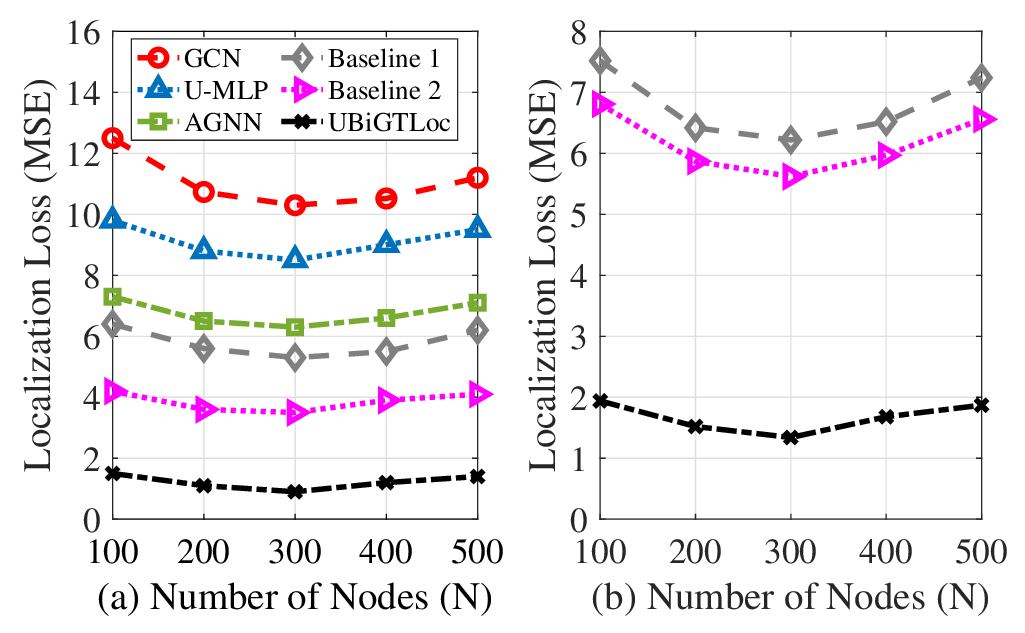}}
    \caption{Localization loss versus the number of sensor nodes 
    in a density-dependent interference environment with $\kappa = 1$ for two different scenarios: (a) anchor-presence ($\alpha = 20\%$), (b) anchor-free ($\alpha = 0$).}
    \label{fig:density effect int}
\end{figure}

\subsection{The Effect of Network Density and Interference on MSE}
In this section, we analyze the impact of network density and interference on localization performance. We investigate two distinct scenarios,  an ideal interference-free scenario with $\kappa = 0$  and a more realistic case that incorporates density-dependent interference with $\kappa = 1$. 
\subsubsection{Performance 
in an Ideal, Interference-Free Environment}
To establish a performance baseline, we first evaluate all methods in an ideal environment where the interference scaling factor, $\kappa$, is set to zero. 
Fig.~\ref{fig:density effect} shows the localization error (i.e., MSE) in meters for different numbers of sensor nodes, ranging from 100 to 500, in the WSN. This experiment evaluates both anchor-presence (i.e., Fig.~\ref{fig:density effect}(a)) and anchor-free (i.e., Fig.~\ref{fig:density effect}(b)) scenarios. In Fig.~\ref{fig:density effect}(a), all methods that rely on the presence of anchor nodes are compared. On the other hand, Fig.~\ref{fig:density effect}(b) focuses solely on methods that can operate in anchor-free environments, including UBiGTLoc and its ablation baselines.

Overall, all methods in both scenarios demonstrate that increasing the number of sensor nodes from 100 to 500 improves the localization error in an ideal, interference-free environment. This is because a higher number of sensor nodes provides more feature information for the model to learn. With more sensor nodes, the model will learn from a greater variety of RSSI information and spatial relationship between sensor nodes. It is further clear that  the proposed UBiGTLoc framework consistently achieves the lowest localization error across all sensor node densities for both scenarios. Notably, the decrease in localization error for UBiGTLoc is also more gradual compared to other methods. For instance, in Fig.~\ref{fig:density effect}(b), while the localization error of Baseline~2 drops from 6.3~m to 1.5~m as the number of sensor nodes increase from 100 to 500, UBiGTLoc only decreases from 1.7~m to 0.5~m. This indicates that UBiGTLoc performs robustly across both sparse and dense network settings thanks to its comprehensive temporal-spatial architecture, which effectively captures intricate patterns in RSSI variations.

\subsubsection{Performance with Density-Dependent Interference}
To evaluate performance under more realistic conditions, we introduce a challenging, high-interference scenario by setting the interference scaling factor to $\kappa = 1$. This experiment investigates the critical trade-off between information gain and interference penalty in dense networks. The results for this more realistic scenario are presented in Fig.~\ref{fig:density effect int}. The corresponding anchor-present scenario is shown in Fig.~\ref{fig:density effect int}(a), and the anchor-free scenario is shown in Fig.~\ref{fig:density effect int}(b). Unlike the ideal, interference-free use case of Fig.~\ref{fig:density effect}, all methods exhibit a U-shaped performance curve under density-dependent interference with $\kappa = 1$. Initially, as the number of sensor nodes increases from 100 to 300, the localization error decreases because the benefit of additional spatial information outweighs the cost of the added interference. However, as the network density increases beyond this point, the trend reverses. In this denser region, the significant interference penalty from each new sensor node surpasses the marginal information gain, causing the overall localization error to rise.
This behavior indicates that the best performance for most methods under the assumed conditions is observed when the network has around 300 sensor nodes, a result specific to the parameters used in our simulation. Notably, even in this challenging, interference-rich environment, UBiGTLoc continues to outperform all other methods in terms of MSE. It further demonstrates greater robustness against interference by providing more stable results, which validates the effectiveness of UBiGTLoc's  comprehensive architecture. This is because it combines  robust temporal feature extraction with adaptive relational reasoning that help in suppressing interference effects compared to existing approaches.



\begin{figure}[t!]
    \centerline{\includegraphics[width=0.5\textwidth]{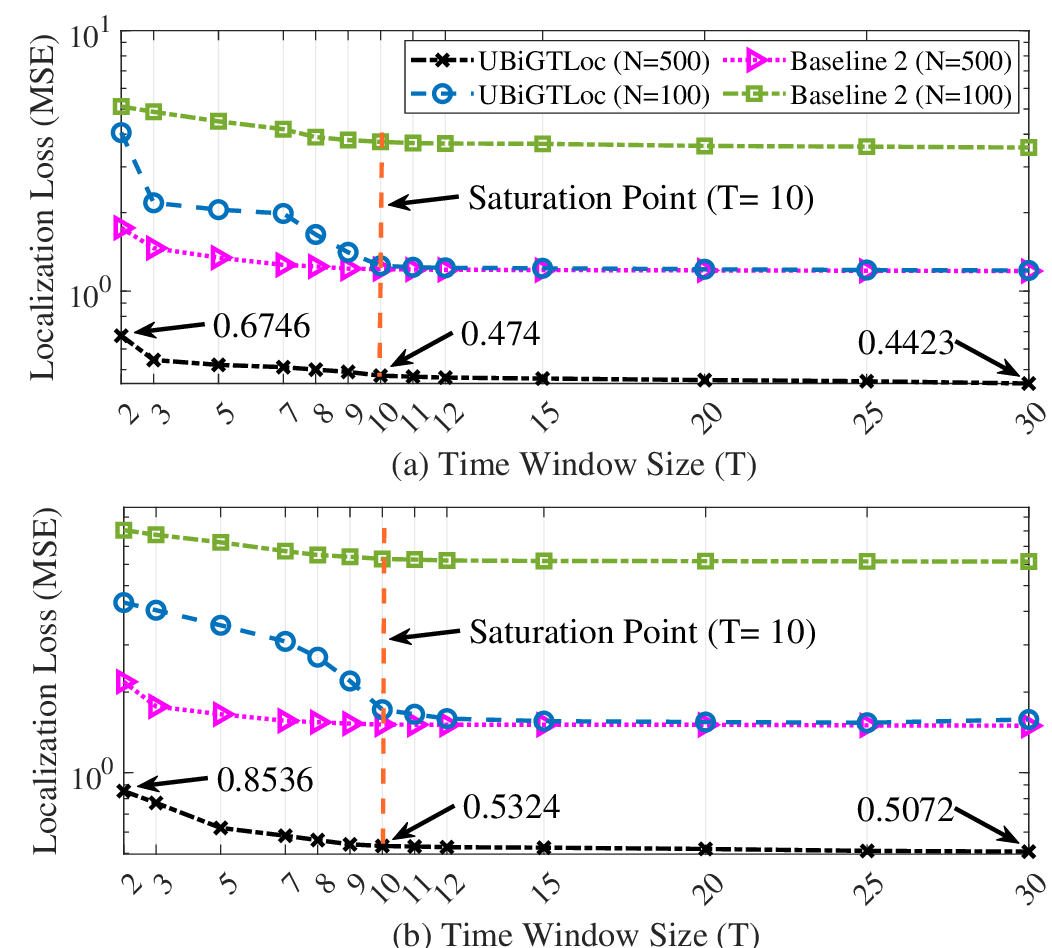}}
    \caption{Localization loss versus time window size with 100 and 500 sensor nodes for two different scenarios: (a) anchor-presence ($\alpha = 20\%$), (b) anchor-free ($\alpha = 0$).}
    \label{fig:T effect}
\end{figure}

 \begin{figure*}[t!]
     \centerline{\includegraphics[width = 0.95\textwidth]{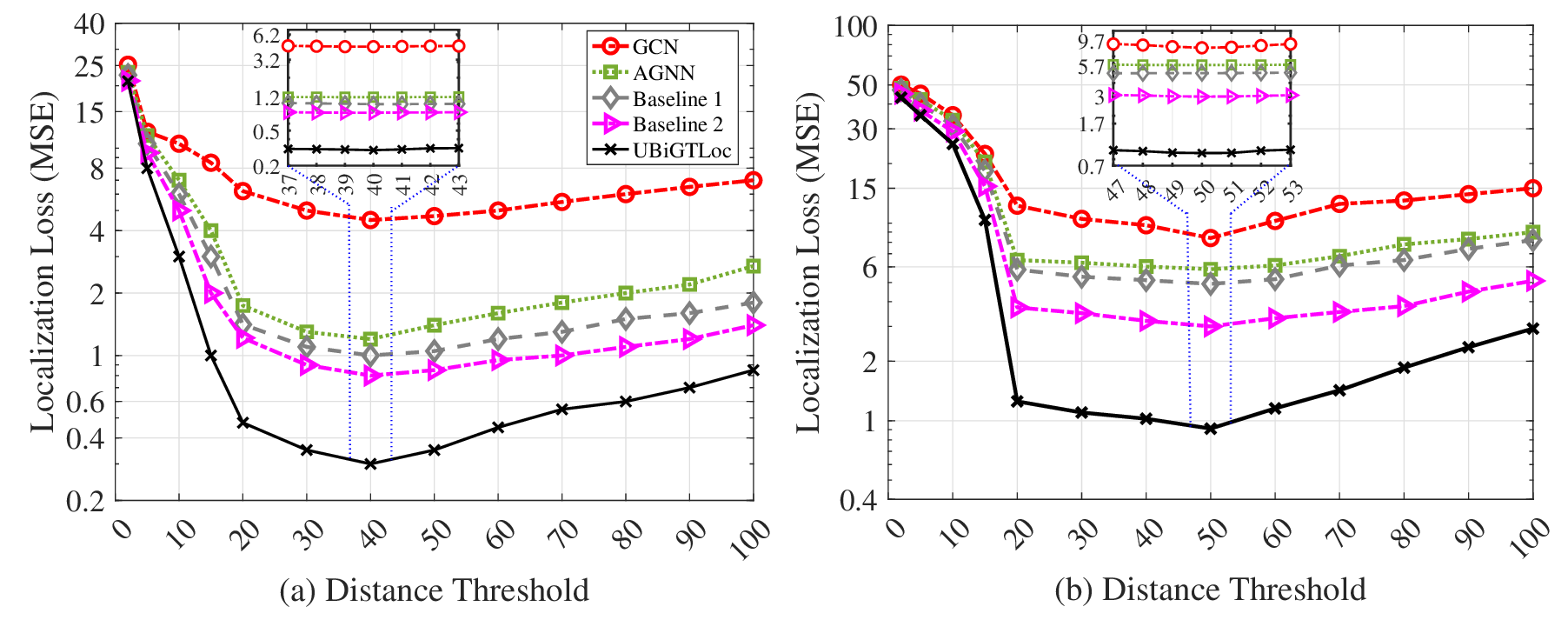}} 
      \caption{Localization loss versus the distance threshold ($d_{th}$) (in meters) for two network settings: (a) 500 sensor nodes, (b) 100 sensor nodes.} 
      \label{Fig: DistanceEffect}
\end{figure*}

\subsection{The Effect of Time Window Size \(T\) on MSE}
In this section, we analyze the impact of the time window size \(T\) on the localization error (i.e., MSE) of UBiGTLoc and Baseline~2 under two different numbers of sensor nodes: 100 and 500. This analysis focuses on these two methods only, as they incorporate temporal information, unlike the other methods.
The results are presented for both anchor-presence  and anchor-free scenarios in Fig.~\ref{fig:T effect}(a) and Fig.~\ref{fig:T effect}(b), respectively. 
Both figures demonstrate that the localization error smoothly decreases with increasing the time window size, which implies incorporating more temporal information. However, Fig.~\ref{fig:T effect}(a) shows a sharp decrease between $2$ and $5$ in UBiGTLoc's localization error in the sparse network setting (i.e., \(N = 100\)). This improvement can be attributed to the reliance on anchor nodes in sparse networks, where the limited number of regular sensor nodes makes anchors critical for improving localization accuracy. A small increase in \(T\) allows UBiGTLoc to effectively leverage the additional temporal information from anchor nodes, resulting in a rapid performance boost. In contrast, such behavior is not observed in a denser network (with $N=500$), because the larger number of regular sensor nodes provides sufficient relative information, diminishing the reliance on anchor nodes. Furthermore, UBiGTLoc consistently performs better than Baseline~2 across all cases. This is primarily due to UBiGTLoc's reliance on BiLSTM, which dynamically captures temporal dependencies, while Baseline~2, which uses EWMA, relies on a fixed weight decay factor. The fixed decay limits Baseline~2's ability to adapt to complex temporal patterns, resulting in its relatively lower performance compared to UBiGTLoc. 
In addition, in both sparse (i.e., \(N = 100\)) and dense (i.e., \(N = 500\)) networks, increasing the time window size \(T\) generally improves the localization performance by allowing the model to capture more temporal information from the RSSI measurements. However, the rate of improvement diminishes beyond a certain point, indicated by the saturation points as shown in Fig.~\ref{fig:T effect}.  
Particularly, the localization error starts to saturate at approximately \(T = 10\) for both scenarios in sparse and dense networks. Increasing the window size beyond this point does not yield further significant improvements, suggesting it as a practical choice for the time window size. For instance, in Fig.~\ref{fig:T effect}(b) with 500 sensor nodes, the MSE for UBiGTLoc decreases from $0.8536$~m at \(T=2\) to $0.5324$~m at \(T=10\), while at \(T=30\), it remains relatively stable at $0.5072$~m, indicating minimal enhancement. Similarly, Baseline~2 shows a decrease from $2.1852$~m at \(T=2\) to $1.5127$~m at \(T=10\), but at \(T=30\), the MSE remains relatively stable at $1.5017$~m. This saturation occurs despite the theoretical expectation that increasing the number of observations should reduce error, as indicated by the Cramér–Rao Bound (CRB)~\cite{10449410}. This is because older RSSI measurements reflect past sensor node locations and movement patterns rather than current position, which reduce their relevance in localization. As temporal correlations weaken, the model’s ability to extract meaningful patterns diminishes, leading to error stabilization and limiting further gains. These practical limitations are aligned with findings in~\cite{10806657}, which discuss how theoretical bounds, like the CRB, can fail to account for the effects of mixed-resolution data on performance. As the study highlights, the CRB assumes ideal conditions, but does not fully capture the complexities of practical conditions.
Generally, selecting a lower time window size~\(T\) improves computational efficiency. In WSNs, sensor nodes need to store the RSSI information in memory and transmit it to a central unit for processing. A smaller time window reduces the amount of data that needs to be stored and transmitted, leading to lower memory requirements and faster processing times. This is particularly important in resource-constrained IoT environments.

\subsection{The Effect of Sensor Nodes Radio Range on MSE}
Fig. \ref{Fig: DistanceEffect} provides results  of how varying the radio range distance threshold (i.e., \(d_{th}\)) impacts the localization error (i.e., MSE) for all methods. The results of U-MLP are omitted since U-MLP is is not a graph-based method. The results are shown for two different network sizes. Specifically, Fig.~\ref{Fig: DistanceEffect}(a) illustrates the case with 500 sensor nodes, while Fig.~\ref{Fig: DistanceEffect}(b) corresponds to the case with 100 sensor nodes. Initially, as the distance threshold increases from 2~m to approximately 20~m, the localization error decreases significantly for all methods.  For instance, in Fig.~\ref{Fig: DistanceEffect}(a), UBiGTLoc sees an improvement from 21.014~m at \(d_{th}=2\) to 0.474~m at \(d_{th}=20\). Similarly, in Fig.~\ref{Fig: DistanceEffect}(b), the localization error for UBiGTLoc drops from 43.1486~m at \(d_{th}=2\) to 1.2523~m at \(d_{th}=20\). This decrease is due to the improved connectivity among the sensor nodes, which allows the model to leverage richer feature information for more accurate sensor node localization. 

For all methods, the localization error continues to decrease between on average 20~m to 40~m as shown in Fig.~\ref{Fig: DistanceEffect}(a), though the rate of improvement decreases with time. Beyond 40~m, the localization error starts to show a relatively slow rise. Similarly, in Fig.~\ref{Fig: DistanceEffect}(b), the localization error decreases between 20~m to 50~m, after which it begins to increase gradually. For instance, GCN’s error rises from 8.4~m at \(d_{th}=50\) to 15~m at \(d_{th}=100\) in Fig.~\ref{Fig: DistanceEffect}(b). This increase is attributed to the introduction of redundant and potentially noisy connections, which occur when the distance threshold is too high. Furthermore, the zoomed-in insets in Fig.~\ref{Fig: DistanceEffect}(a) and Fig.~\ref{Fig: DistanceEffect}(b) provide a granular view of performance trends around the critical thresholds. Within these narrow ranges, localization error stabilizes near 40~m and 50~m, respectively, before rising steadily as connectivity thresholds increase. This emphasizes the need to manage connectivity near 40~m and 50~m to avoid noise-induced performance degradation. In real-world scenarios, this represents the radio range within which sensors can communicate. Excessive connectivity complicates the network structure and adds noise, ultimately impairing the model’s accuracy. Therefore, all methods exhibit best performance within the radio range of 20 to 40~m in WSNs of 500 sensor nodes and 20 to 50~m in WSNs of 100 sensor~nodes.

\begin{figure}[t!]
     \centerline{\includegraphics[width = 0.5\textwidth]{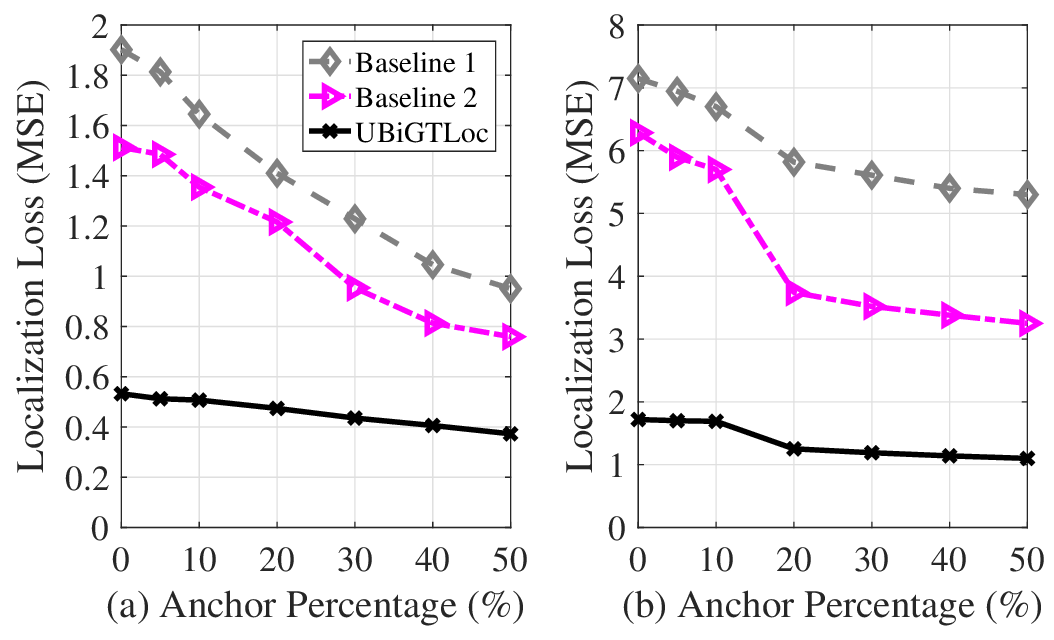}} 
      \caption{Localization loss versus anchor nodes percentage (\(\alpha\)) for two network settings: (a) 500 sensor nodes, (b) 100 sensor nodes.} 
     
      \label{anchor-percentage-effect}
\end{figure}

\subsection{The Effect of Anchor Nodes Percentage on MSE} \label{subsection: anchor_percentage}
Fig.~\ref{anchor-percentage-effect} illustrates the impact of varying anchor node percentages (\(i.e., \alpha\)) on localization error (i.e., MSE) for Baseline~1, Baseline~2, and UBiGTLoc. These methods are selected as they operate in both anchor-presence and anchor-free environments. Results are shown for two network sizes. Particularly, Fig.~\ref{anchor-percentage-effect}(a) shows the results for 500 sensor nodes, and Fig.~\ref{anchor-percentage-effect}(b) shows the results for 100 sensor nodes.  
The results indicate that increasing the percentage of anchor nodes leads to a general decrease in localization error across all methods. This is expected because anchor nodes provide stable spatial references to enhance localization performance. UBiGTLoc, on the other hand, shows a slower but steady decrease in localization error as anchor node density increases. 
Even with fewer anchor nodes, UBiGTLoc maintains a significant performance advantage due to its superior handling of spatial-temporal information. In Fig.~\ref{anchor-percentage-effect}(a), UBiGTLoc achieves an MSE of 0.5324~m in the zero-anchor scenario, which decreases to 0.3728~m when 50\% of the sensor nodes are anchor nodes. In contrast, Baseline~1 experiences a more significant drop in MSE from 1.9017~m with 0\% anchor nodes to 0.9506~m with 50\% anchor nodes. Baseline~2 shows a reduction from 1.5127~m to 0.7594~m under the same conditions. This indicates that UBiGTLoc maintains a more stable performance even with fewer anchor nodes.
Similarly, in Fig.~\ref{anchor-percentage-effect}(b), UBiGTLoc starts with an MSE of 1.7191~m without anchor nodes and drops to 1.1~m with 50\% anchor nodes. On the other hand, Baseline~1 and Baseline~2 show greater reliance on anchor nodes, with Baseline~1 going from 7.1483~m to 5.3~m and Baseline~2 from 6.2847~m to 3.25~m, respectively.
The results highlight that UBiGTLoc achieves a steady decrease in localization error as anchor density increases. It remains effective even in anchor-free scenarios, unlike other methods which show a sharper dependency on the presence of anchor nodes for improved accuracy.

\subsection{Discussion of Spatiotemporal WSN Localization Challenges}
The results presented above collectively demonstrate the robustness of the proposed UBiGTLoc framework, which outperforms the traditional anchor-based sensor nodes' localization methods. Effectively capturing spatiotemporal relationships in WSNs poses multiple non-trivial challenges, which UBiGTLoc addresses through careful design and fine-tuning. 
\subsubsection{Scalability in Large-Scale WSNs} Scalability is  an essential  concern in large WSNs. Conventionally, increasing the number of sensor nodes and connections can lead to higher computational complexity in both BiLSTM and graph-based models. However, since each sensor node in UBiGTLoc processes only its neighboring sensor nodes’ data, the computation remains localized. Thus, UBiGTLoc framework maintains steady operation across varying network densities (i.e., $100-500$ sensor nodes per WSN field) without incurring excessive computational complexity.
\subsubsection{Fine-Tuning the Temporal Window for Balanced Performance} Fine-tuning the temporal window size (i.e., \(T\)) is crucial as a small window may fail to capture meaningful trends, while a large one can unnecessarily increase complexity. UBiGTLoc carefully balances this trade-off, using BiLSTM-based temporal encoding which surpasses other temporal-based encoding solutions (such as EWMA), achieving excellent performance with a reasonable window size of $10$~timestamps while maintaining performance efficiency.
\subsubsection{Enhancing Generalization with Limited Data} A major challenge in WSN localization is the limited availability of labeled data, which can lead to overfitting. To mitigate this, UBiGTLoc incorporates 
 data augmentation methods to preserve the spatiotemporal structure of RSSI measurements. This enhances its ability to learn spatiotemporal patterns effectively, ensuring reliable localization even in unseen WSN~topologies.
\subsubsection{Architectural Components  and Hyperparameter Tuning} Properly configuring the model architecture is crucial for achieving generalization across different unseen WSN topologies. We optimize key components, including BiLSTM hidden units, the number of graph transformer layers, dropout rates, and batch normalization. Additionally, by using only the final hidden states from each BiLSTM direction, we condense sequence information into a compact representation, reducing dimensionality and complexity while maintaining contextual integrity. These design choices ensure UBiGTLoc remains efficient and adaptable across diverse network conditions.

 Despite its strengths, there are challenges to be addressed in the future. For instance, the performance of UBiGTLoc in even sparser networks needs further investigation. Additionally, the centralized nature of UBiGTLoc presents drawbacks, such as increased energy consumption and privacy concerns related to the collection and sharing of RSSI data with the central node. These issues could be mitigated through future distributed implementations of UBiGTLoc.

\section{Conclusion} \label{sec:conclusion}
In this article, we proposed  UBiGTLoc, a unified sensor nodes' localization framework for wireless IoT sensor networks. In contrast to existing methods, the proposed UBiGTLoc framework is capable of operating in both anchor-free and anchor-presence WSNs. For this purpose, it utilizes the  RSSI data from each sensor node and the adjacency matrix to capture the temporal and spacial variations, respectively, for accurate sensor nodes' localization. 
Along with this, UBiGTLoc demonstrates superior performance compared to the considered existing methods such as GCN, U-MLP, and AGNN and the proposed two baselines. It excels in accuracy, adapts well in dynamic IoT environments, and remains robust in noisy conditions. Additionally, the proposed UBiGTLoc framework  operates effectively in both sparse and dense networks, providing a versatile and cost-effective solution for various IoT applications.

In future research, we plan to explore multi-modal data integration with UBiGTLoc by incorporating imagery from ground-based cameras, drones, or satellite systems. By combining visual data with RSSI measurements, the model can leverage environmental context and spatial features to further enhance localization accuracy, especially in challenging IoT environments, such as those of remote or complex urban settings. Additionally, we plan to investigate collaborative localization with edge computing. This can decentralize the computation process by enabling sensor nodes to work together in determining their positions. This approach can further alleviate the load on the central unit, improve real-time processing, and enhance system responsiveness for latency-sensitive applications.




\bibliography{bibtex/IEEEexample,bibtex/IEEEabrv,bibtex/UBiGLoc}

\end{document}